\pdfoutput=1
\documentclass[twocolumn]{aastex62}

\usepackage{xspace}

\newcommand{\Mgas}{\ensuremath{M_{\rm{H_{2}}}}\xspace}
\newcommand{\Mstar}{\ensuremath{M_{\rm{*}}}\xspace}

\newcommand{\Mhalo}{\ensuremath{M_{\rm{halo}}}\xspace}
\newcommand{\fgas}{\ensuremath{f_{\rm{H_{2}}}}\xspace}
\newcommand{\tdep}{\ensuremath{t_{\rm{dep}}}\xspace}

\newcommand{\msol}{\ensuremath{\rm{M}_\odot}\xspace}

\newcommand{\lprime}{\ensuremath{\rm{L}_{\rm{CO}}'}\xspace}

\newcommand{\alphaco}{\ensuremath{\alpha_{\rm{CO}}}\xspace}

\graphicspath{{./}{figures/}}

\shortauthors{Williams et al.}

\begin{document}

\title{ALMA measures rapidly depleted molecular gas reservoirs in massive quiescent galaxies at z$\sim$1.5}

\author[0000-0003-2919-7495]{Christina C. Williams}\altaffiliation{NSF Fellow}
\affiliation{Steward Observatory, University of Arizona, 933 North Cherry Avenue, Tucson, AZ 85721, USA}

\author[0000-0003-3256-5615]{Justin~S.~Spilker}
\altaffiliation{NHFP Hubble Fellow}
\affiliation{Department of Astronomy, University of Texas at Austin, 2515 Speedway, Stop C1400, Austin, TX 78712, USA}

\author[0000-0001-7160-3632]{Katherine E. Whitaker}
\affiliation{Department of Astronomy, University of Massachusetts, Amherst, 710 N. Pleasant Street, Amherst, MA 01003, USA}
\affiliation{Cosmic Dawn Center (DAWN)}

\author{Romeel Dav\'e}
\affiliation{Institute for Astronomy, Royal Observatory, Edinburgh EH9 3HJ, United Kingdom}

\author{Charity Woodrum}
\affiliation{Steward Observatory, University of Arizona, 933 North Cherry Avenue, Tucson, AZ 85721, USA}

\author{Gabriel Brammer}
\affiliation{Niels Bohr Institute, University of Copenhagen, Lyngbyvej 2, DK-2100 Copenhagen, Denmark}
\affiliation{Cosmic Dawn Center (DAWN)}

\author[0000-0001-5063-8254]{Rachel Bezanson}
\affiliation{Department of Physics and Astronomy and PITT PACC, University of Pittsburgh, Pittsburgh, PA, 15260, USA}

\author[0000-0002-7064-4309]{Desika Narayanan}
\affiliation{Department of Astronomy, University of Florida, 211 Bryant Space Science Center, Gainesville, FL 32611, USA}
\affiliation{Cosmic Dawn Center (DAWN)}

\author[0000-0001-6065-7483]{Benjamin Weiner}
\affiliation{Steward Observatory, University of Arizona, 933 North Cherry Avenue, Tucson, AZ 85721, USA}

\begin{abstract}
We present ALMA CO(2--1) spectroscopy of 6 massive (log$_{10}$\Mstar/\msol$>$11.3) quiescent galaxies at $z\sim1.5$. These data represent the largest sample 
using CO emission to trace molecular gas 
in quiescent galaxies above $z>1$, achieving an average 3$\sigma$ sensitivity of \Mgas$\sim10^{10}$\msol. 
We detect one galaxy at 4$\sigma$ significance and place upper limits on the molecular gas reservoirs of the other 5, finding molecular gas mass~fractions $\Mgas/\Mstar=\fgas<2-6\%$ (3$\sigma$ upper limits). This is 1-2 orders of magnitude lower than coeval star-forming galaxies at similar stellar~mass, 
and comparable to galaxies at $z=0$ with similarly low sSFR. This indicates that their molecular gas reservoirs were rapidly and efficiently used up or destroyed, and that gas fractions are uniformly low ($<$6\%) despite the structural diversity of our sample. The implied rapid depletion time of molecular gas (\tdep$<0.6$ Gyr) disagrees with extrapolations of empirical scaling relations to low sSFR. We find that our low gas fractions are instead in agreement with predictions from both the recent \textsc{simba} cosmological simulation, and from analytical ``bathtub" models for gas accretion onto galaxies in massive dark~matter~halos (log$_{10}\Mhalo/\msol\sim14$  at $z=0$). 
Such high~mass halos reach a critical mass of log$_{10}\Mhalo/\msol>12$ by $z\sim4$ that halt the accretion of baryons early in the Universe. 
Our data is consistent with a simple picture where galaxies truncate accretion and then consume the existing gas at or faster than typical main sequence rates. Alternatively, we cannot rule out that these  galaxies reside in lower~mass halos, and low gas fractions may instead reflect either stronger feedback, or more efficient gas consumption.

\end{abstract}

\section{Introduction} \label{sec:intro}

A challenge of modern galaxy evolution is to understand the formation of massive and quiescent galaxies. Stellar archaeology indicates that massive galaxies (log$_{10}$\Mstar/\msol$>$11) form their stars in a rapid burst in the first 1-3 billion years of the universe ($z>2$) \citep[e.g.][]{Thomas2010, McDermid2015}.
After this rapid growth phase, their star formation halted (quenched) through unknown processes, and most remain dormant, without significant star formation for $>$10 billion years \citep[e.g.][]{Renzini2006,Citro2016}.

The observed rapid growth and early death in quenched galaxies are longstanding problems in our theoretical understanding of galaxy formation. This is particularly true for high mass galaxies at high redshifts, which have caused the largest tension with simulations in terms of reproducing numbers \citep[e.g.][]{Santini2012,Cecchi2019} and halting and preventing further star formation \citep[e.g.][]{Croton2006, NaabOstriker2017, Forrest2020}. 
Recent improvements to the physical models that are implemented in cosmological simulations are well matched to the observed properties of massive quiescent galaxies at least from $z<2.5$, the growth of their black holes, and the maintenance of quiescence across cosmic time \citep[e.g.][]{Schaye2015,Feldmann2017, Nelson2018, Pillepich2018}. However, even with recent advances, simulations generally require some form of poorly understood, yet extreme feedback to truncate star formation and reproduce the properties of observed massive galaxies across cosmic time \citep[for a review see][]{SomervilleDave2015}.

A key unknown is the evolution of the cold gas reservoirs, the fuel for star formation, in massive galaxies as they transition from star forming to quiescent. While the majority of massive galaxies quench around cosmic noon \citep[$1<z<3$;][]{Whitaker2011, Muzzin2013mf, Tomczak2014,Davidzon2017}, surveys characterizing the molecular gas reservoirs using rotational transitions of CO generally find that cold gas is abundant in massive star forming galaxies during this era. The continuity of the star forming sequence implies high accretion rates from the intergalactic medium \citep[for reviews, see][]{Tacconi2020, HodgeDaCunha2020}.

To quench, galaxies must break this equilibrium. To sufficiently deplete, expel, or heat the
abundant gas supply in massive galaxies, theoretical quenching
models favor strong feedback from  supermassive black holes \citep[]{Choi2017, Weinberger2017, Weinberger2018} 
or extreme star formation \citep{Hopkins2010,  Grudic2019}. 
These may be driven by efficient and rapid growth \citep{Wellons2015, Williams2014, Williams2015}, mergers \citep{diMatteo2005,Hopkins2006} or disk-instabilities \citep{DekelBurkert2014,Zolotov2015}. 
Additional theories exist to stabilize existing cold gas from collapse, e.g. through the growth of a stellar bulge \citep{Martig2009}, thereby decreasing the star formation efficiency to quench. However, this likely requires that accretion be halted \citep[through shock-heating at the virial radius for massive dark matter halos with log$_{10}$\Mhalo/\msol$>$12; e.g.][]{BirnboimDekel2003,Keres2005, DekelBirnboim2006}. 
To first order, these different mechanisms (destruction by feedback, consumption, or stabilization) yield different predictions for the rate at which cold gas disappears from galaxies relative to ceasing star formation.

The evolution of molecular gas reservoirs in quiescent galaxies is therefore an important constraint on the possible mechanisms halting star formation. 
Several surveys have characterized molecular gas in massive quiescent galaxies using CO at $z\sim0$ \citep[][]{Young2011, Saintonge2011, Saintonge2012, Saintonge2017,Davis2016}, generally finding that galaxies maintain low gas fractions ($<$0.1-1\%) after $\gtrsim$10 Gyr of quiescence. However, the peak epoch of the transition to quiescence for massive galaxies is at $z\sim2$, where few observations have been made to date.

Characterizing the distribution of molecular gas reservoirs in quenched galaxies would be a major step forward in understanding the pathways massive galaxies take to quiescence. With this work, we conduct the first survey targeting molecular gas traced by CO(2--1) in a sample of quiescent galaxies above $z>1$. These build on samples studied at $z\sim0$ \citep{Rowlands2015, French2015, Alatalo2016}, at intermediate redshifts \citep{Spilker2018, Suess2017}, and at $z>1$, single galaxies \citep{Sargent2015, Bezanson2019}, and average properties through stacking dust emission \citep{Gobat2018}. From these studies emerges a wide diversity of molecular gas properties in quenched and quenching galaxies. Key informative constraints on the molecular gas reservoirs include 1) their variation with properties related to quiescence, such as compact stellar density \citep[e.g.][]{Whitaker2017, Lee2018},  in light of recent reports that age and quenching timescale varies with size and stellar density \citep{Williams2017,Wu2018,Belli2019}
and 2) the amount of gas leftover relative to the time galaxies stopped forming stars, tracing the timescale for consumption.

In this paper, we present a new survey with the Atacama Large Millimeter/submillimeter Array (ALMA) targeting the CO(2--1) emission in quiescent galaxies at $z>1$. In Section \ref{sec:data} we present our sample, their stellar population properties, and our ALMA observations of the CO(2--1) emission line. 
In Section \ref{sec:results}, we present our ALMA measurements in the context of other molecular gas surveys, and in Section \ref{sec:discussion}, we discuss our results in the context of theoretical ideas about the formation of quiescent galaxies. We assume a $\Lambda$CDM cosmology with  H$_0$=70 km s$^{-1}$ Mpc$^{-1}$, $\Omega_M$ = 0.3, $\Omega_\Lambda$ = 0.7, and a \citet{Chabrier2003} initial mass function (IMF).

\section{Sample and Data}\label{sec:data}

\begin{figure*}[th]
\includegraphics[scale=.58, trim=50 50 50 50,clip]{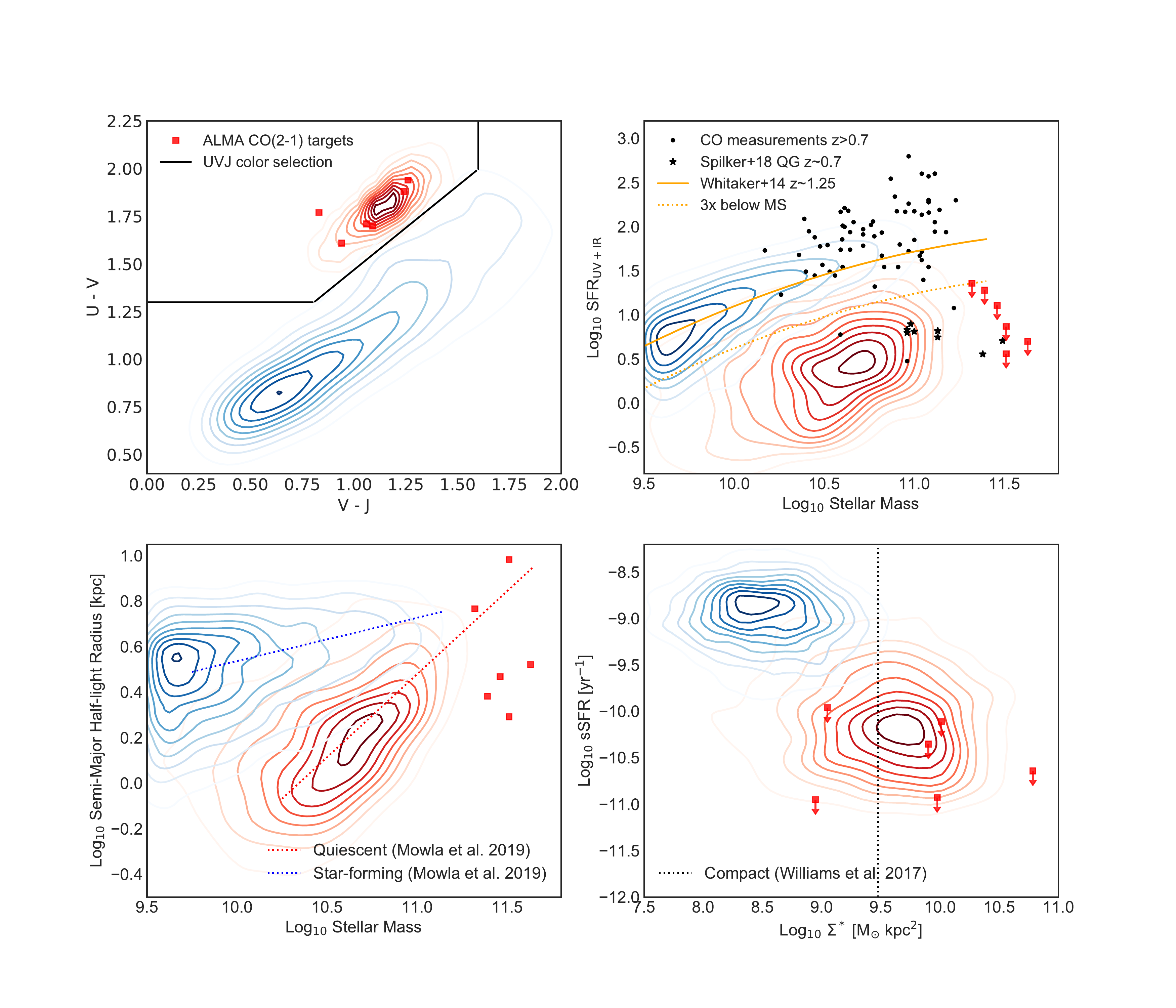}
\caption{ Our ALMA targets (red squares) compared to star-forming and quiescent galaxies from 3DHST with log$_{10}$\Mstar/\msol$>9.5$ at $1<z<1.5$ \citep[blue/red contours][]{Skelton2014}. 
Top Left: U-V vs V-J rest-frame colors and quiescent galaxy selection (black). 
Top Right: Star formation rate vs stellar mass.  Our sample are more than 3$\times$ below the main sequence at $1<z<1.5$ \citep[][]{Whitaker2014}. Black points show CO observations at $z>0.7$ for star forming galaxies (see Section \ref{sec:results}).
Bottom Left: Size vs mass distributions and mean relations at $z\sim1.25$ \citep{Mowla2019}.
Bottom Right: sSFR vs stellar surface density, $\Sigma_* = \Mstar/2\pi R_{e}^2$.  
Stellar density higher than the dotted line indicates quiescent galaxies that are compact as defined in \citep{Cassata2013, Williams2017}. Bottom panels demonstrate that our ALMA sample spans the full range of quiescent galaxy sizes and densities at this redshift. }
\label{fig:props}
\end{figure*}

We select ALMA targets from the literature of spectroscopically confirmed quiescent galaxies at redshifts $1<z<1.74$, where the CO(2--1) molecular transition is observable in ALMA Band 3, and also have state-of-the-art ancillary data (deep rest-frame UV to mid-IR coverage, including high-resolution {\it Hubble Space Telescope (HST)} WFC3 imaging).
We identified the most massive of those published (log$_{10}\Mstar/\msol>11.3$) that have quiescent stellar populations based on both their rest frame optical spectroscopy, UV-VJ rest-frame colors (Figure \ref{fig:props}) and UV+IR star formation rates. 
Our final sample includes five galaxies in the COSMOS field, which are all confirmed to be quiescent on the basis of deep Balmer absorption features, strong Dn4000, and a lack of strong emission lines, using deep spectroscopy from Keck/LRIS \citep{Bezanson2013,vandeSande2013, Belli2014a, Belli2015} and Subaru/MOIRCS \citep{Onodera2012}. In this study, we combine these five targets with one galaxy from our pilot observation published in \citet{Bezanson2019}.

\subsection{Optical and infrared data}
The five galaxies confirmed with Keck were originally selected for spectroscopy from the NEWFIRM Medium Band Survey \citep[NMBS;][]{Whitaker2011}. NMBS includes multi-wavelength photometry from the UV to 24 $\mu$m, and in particular, medium-band near-IR filters that sample the Balmer/4000 \AA\ break at our target redshifts. The Subaru target from the sample of \citet{Onodera2012} was selected from the BzK color-selected catalog published in \citet{Mccracken2010}.

To identify our target galaxies, we used the stellar masses measured from the photometric spectral energy distribution (SED) as published in the original studies. These works fit the photometry using FAST \citep{Kriek2009} to estimate stellar masses assuming \citet[][]{Bruzual2003} stellar population models with exponentially declining star formation history (SFH), a \citet[][]{Chabrier2003} IMF and \citet{Calzetti2000} dust attenuation. Where relevant, we convert literature measurements to Chabrier IMF for comparison to our measurements.  Not all targets had stellar ages measured in the literature (based on the UV to IR photometry), but where available they indicate old stellar ages (1-1.5 Gyr), with the exception of our pilot galaxy 21434, whose published stellar age is 800 Myr \citep{Bezanson2019}. The pilot galaxy's rest-frame colors are also the closest to the bluer post-starburst region of the UVJ-quiescent diagram \citep{Whitaker2012,Belli2019}. 

For results presented herein, we re-fit the UV to near-IR photometry uniformly using the SED-fitting code \textsc{prospector} \citep{Johnson2019}, which uses the Flexible Stellar Populations Synthesis (FSPS) code \citep{Conroy2009,ConroyGunn2010}. We fit using the \textsc{prospector}-$\alpha$ model framework \citep{Leja2017} which includes a non-parametric SFH that has been shown to be more realistic and physically representative of massive galaxies \citep{Leja2019a}. For the purpose of fitting the stellar population properties probed by the UV to near-IR photometry, we augment the \textsc{prospector}-$\alpha$ model by removing emission due to active galactic nuclei (AGN; which contribute primarily at mid-IR wavelengths) and the dust emission.
Re-fitting the galaxies uniformly in this way also enables us to measure the mass-weighted age, which is more directly comparable to the cosmological simulations we present in Section \ref{sec:discussion}. 

We use the NMBS photometric catalog  
that includes medium band near-infrared photometry for all galaxies, with the exception of one (ID 307881) which lies outside the NMBS footprint. For this galaxy we use the UltraVISTA catalog with broad-band photometry in the near-infrared \citep{Muzzin2013}. We present the stellar population properties measured with \textsc{prospector} using the modified \textsc{prospector}-$\alpha$ model and default priors in Table \ref{tab:sedfit}.  Using these fits instead of the literature values results in an average difference of $\sim0.1$dex higher stellar mass. We find a similar difference in stellar masses 
between using a non-parametric SFH and an exponentially declining SFH within \textsc{prospector}. This difference in mass with assumed SFH is consistent with that characterized for massive log$_{10}$\Mstar/\msol$>$11 galaxies \citep{Leja2019b}. Our results do not significantly depend on the choice of SFH or its impact on measured stellar mass, which affect our measurements of \fgas\ by less than a factor of 1.5.

A second impact of the non-parametric SFH is that the mass-weighted age of the galaxies are typically older than that derived using parametric SFH \citep{Leja2019b}. While the ages measured assuming an exponentially declining model are typically of order 1-3 Gyr, the non-parametric model returns ages of order 2-3 Gyr. These imply that the major star formation episodes in our sample happened above $z>3$. We list both values in Table \ref{tab:sedfit}. In the rest of this work, stellar age will refer to mass-weighted age, and we adopt the older ages from the non-parametric model because it is the more conservative constraint, as we will discuss in Section \ref{sec:timescale}. 

\subsubsection{Estimation of the star formation rates}\label{sec:sfr}
In this work we consider star formation rate (SFR) measured using two different methods, from the SED fitting outlined in the last section, and also that measured by modeling the obscured and unobscured fluxes, SFR$_{\rm UV+IR}=$SFR$_{\rm UV,uncorr}+$SFR$_{\rm IR}$ as  published in the UltraVISTA catalog \citep{Muzzin2013}. The SFR$_{\rm UV,uncorr}$ is calculated using the conversion of \citet[][]{Kennicutt1998} and the IR component is extrapolated from observed 24$\mu$m flux following \citet[][]{Wuyts2008}. 
SFRs from either method of SED fitting or extrapolated from 24$\mu$m are uncertain for quiescent galaxies. In particular, the SFR$_{\rm UV+IR}$ should be considered an upper limit, because of significant contributions to the mid and far infrared flux that do not trace ongoing star formation in older galaxies \citep[e.g. asymptotic giant branch (AGB) stars, AGN, dust heated by older stars][]{Salim2009, Fumagalli2014, Utomo2014, Hayward2014}. We list the SFR measured using both indicators in Table \ref{tab:sedfit}, and in the rest of this work we adopt SFR$_{\rm UV+IR}$ when measuring sSFR, which we explicitly consider to be an upper limit. Our upper limits to the sample sSFR range from $-10< \rm Log_{10} sSFR < -12$ yrs$^{-1}$.

\begin{deluxetable*}{lccccccccccc}[t!]
\tablecaption{Properties of ALMA targets } 
\tablecolumns{10}
\tablewidth{0pt}
\tablehead{
\colhead{ID$^a$} &
\colhead{RA} &
\colhead{Dec} &
\colhead{$z_{spec}$} &
\colhead{Mass} &
\colhead{SFR$_{\rm UV+IR}$$^b$} &\colhead{SFR$_{\rm 30Myr}$$^c$} & \colhead{Re[kpc]$^d$} & \colhead{Age$^e$} & \colhead{Age$^f$} &\colhead{Reference} }
\startdata
22260 & 149.818229 & 2.561610 & 1.240 & 11.51 $^{+ 0.04 }_{- 0.03 }$ &  3.6 & 5.3 $^{+ 3.41 }_{- 1.91 }$ &  7.6 & 3.4 &  4.6 & Bezanson+2013 \\
20866 & 149.800931 & 2.537990 & 1.522 & 11.46 $^{+ 0.03 }_{- 0.03 }$ &  12.8 & 0.7 $^{+ 2.69 }_{- 0.68 }$ &  2.4 & 2.4 &  1.7 & Bezanson+2013 \\
34879 & 150.131380 & 2.523800 & 1.322 & 11.32 $^{+ 0.04 }_{- 0.04 }$ &  22.9 & 1.4 $^{+ 2.30 }_{- 1.20 }$ &  5.5 & 2.5 &  2.1 & Belli+2015 \\
34265 & 150.170160 & 2.481100 & 1.582 & 11.51 $^{+ 0.03 }_{- 0.03 }$ &  7.4 & 0.3 $^{+ 1.61 }_{- 0.34 }$ &  0.9 & 2.1 &  1.3 & Belli+2015 \\
21434 & 149.816230 & 2.549250 & 1.522 & 11.39 $^{+ 0.03 }_{- 0.03 }$ &  19.1 & 0.5 $^{+ 1.79 }_{- 0.49 }$ &  1.9 & 2.1 &  1.2 & Bezanson+2013,2019 \\
307881 & 150.648487 & 2.153990 & 1.429 & 11.63 $^{+ 0.03 }_{- 0.03 }$ &  5.0 & 0.7 $^{+ 1.73 }_{- 0.66 }$ &  2.7 & 2.7 &  3.2 & Onodera+2012 \\
\enddata \label{tab:sedfit}
\tablenotetext{a}{We adopt IDs as published in the source reference. ID for galaxy 34265 is from \cite{Belli2015} but is referred to as NMBS-COSMOS18265 in \citet{vandeSande2013}. }
\tablenotetext{b}{SFR$_{\rm UV+IR}$ values correspond to those published by the ULTRAVISTA survey \citet[][]{Muzzin2013}.}
\tablenotetext{c}{Corresponds to the average SFR over the past 30 Myr as derived from our SED fitting with \textsc{prospector}.}
\tablenotetext{d}{Circularized half light radius, defined as $R_{e} = r_{e}\sqrt{b/a}$ where $r_{e}$ is the semi-major  axis and $b/a$ is  the  axis  ratio. Measured with GALFIT \citep{Peng2002}}
\tablenotetext{e}{Mass-weighted stellar age as derived from fitting with non-parametric SFH \textsc{prospector} (in Gyr).}
\tablenotetext{f}{Mass-weighted stellar age assuming an exponentially declining SFH (in Gyr).}
\end{deluxetable*}

\subsubsection{Hubble Space Telescope imaging}

Our ALMA target selection includes the requirement of high-resolution rest-frame optical imaging from {\it HST} to enable accurate measurements of morphology of the sample. Because compactness is known to be the strongest predictor of quiescence \citep{Franx2008,Bell2012, Teimoorinia2016,Whitaker2017} this requirement enables an assessment of a possible additional correlation with gas content. Our selected galaxies are structurally representative for their redshift, spanning a large range of half-light radius ($R_e$) and stellar densities ($\Sigma_{\star}\propto M_{\star}$/Re$^{2}$) among quiescent galaxies (Figure \ref{fig:props}).

We process all available {\it HST} imaging covering the ALMA sources with the {\sc grizli} software\footnote{https://github.com/gbrammer/grizli} (Brammer, in prep.). These include WFC3/F160W imaging from programs 12167, 14114, 12440, and ACS F814W imaging from programs 10092 and 9822.  Briefly, we first group all exposures into associations defined as exposures taken with a single combination of instrument, bandpass filter, and guide-star acquisition (i.e., a ``visit'' in the standard {\it Hubble} nomenclature).  We align all individual exposures in an association to each other allowing small shifts to the original astrometry from the files downloaded from the MAST archive at STScI.  For the global astrometry, we generate a reference astrometric catalog from sources in the ultra-deep optical catalog of the entire COSMOS field provided by the HyperSurprime-Cam Subaru Strategic Program \citep[DR2;][]{Aihara2019}, 
which we have verified is well aligned to the GAIA DR2 reference frame \citep{GAIA2016,GAIA2018}. 
We align the {\it HST} association exposures as a group to this reference catalog, allowing for corrections in shift, scale and rotation, resulting in a final global astrometric precision of $\sim$30\,milli-arcseconds. Finally, we combine exposures in a given filter (from one or more associations) using {\sc DrizzlePac} / {\sc AstroDrizzle} \citep{Gonzaga2012}.

\subsection{ALMA observations}
The ALMA observations of our target galaxies were carried out in project 2018.1.01739.S (PI: Williams) in separate observing sessions from December 18, 2018 to January 17, 2019 using the Band 3 (3\,mm) receivers. We combine the results from this program with ALMA data for one similar galaxy from a previous pilot program \citep[2015.1.00853.S; see][for details]{Bezanson2019}. 

The correlator was configured to center the CO(2--1) line within a spectral window of 1.875\,GHz width, which provides $\sim$5500\,km\,s$^{-1}$ of bandwidth centered on the expected frequency of the CO line, $\approx$89.3--102.9\,GHz. Three additional spectral windows were used for continuum observations. Targets 20866, 22260, and 307881 were observed for a total of $\sim$90--100\,min on-source, while 34265 and 34879 were each observed for about twice as long. The array was in a compact configuration yielding synthesized beam sizes $\sim$1.5--2.5'' so as not to spatially resolve the target galaxies. Bandpass and flux calibrations were performed using J1058+0133 and gain calibration using J0948+0022. The data were reduced using the standard ALMA pipeline and the reductions checked manually. Our cleaning procedure involved first masking regions with clearly-detected emission (S/N $>$ 5) and then we used a stopping criterion of 3$\times$ the image rms.

Images of both the continuum and line emission were created using natural weighting of the visibilities to maximize sensitivity, with pixel sizes chosen to yield 5--10 pixels across the synthesized beam. The spectral cubes have a typical noise of 50-65$\mu$Jy/beam in a 400km/s channel measured near the rest-frequency of the CO(2--1) line. The continuum data combined all available spectral windows, and reach a typical sensitivity of $\sim5-9\mu$Jy/beam, calculated as the rms of the non-primary-beam corrected image. All target galaxies are undetected in the continuum. However, the continuum imaging yielded several serendipitous 3-mm sources in these deep data. Two of these continuum sources were previously unknown galaxies and are presented in \citet{Williams2019}.

\begin{figure*}[]
\includegraphics[scale=0.6]{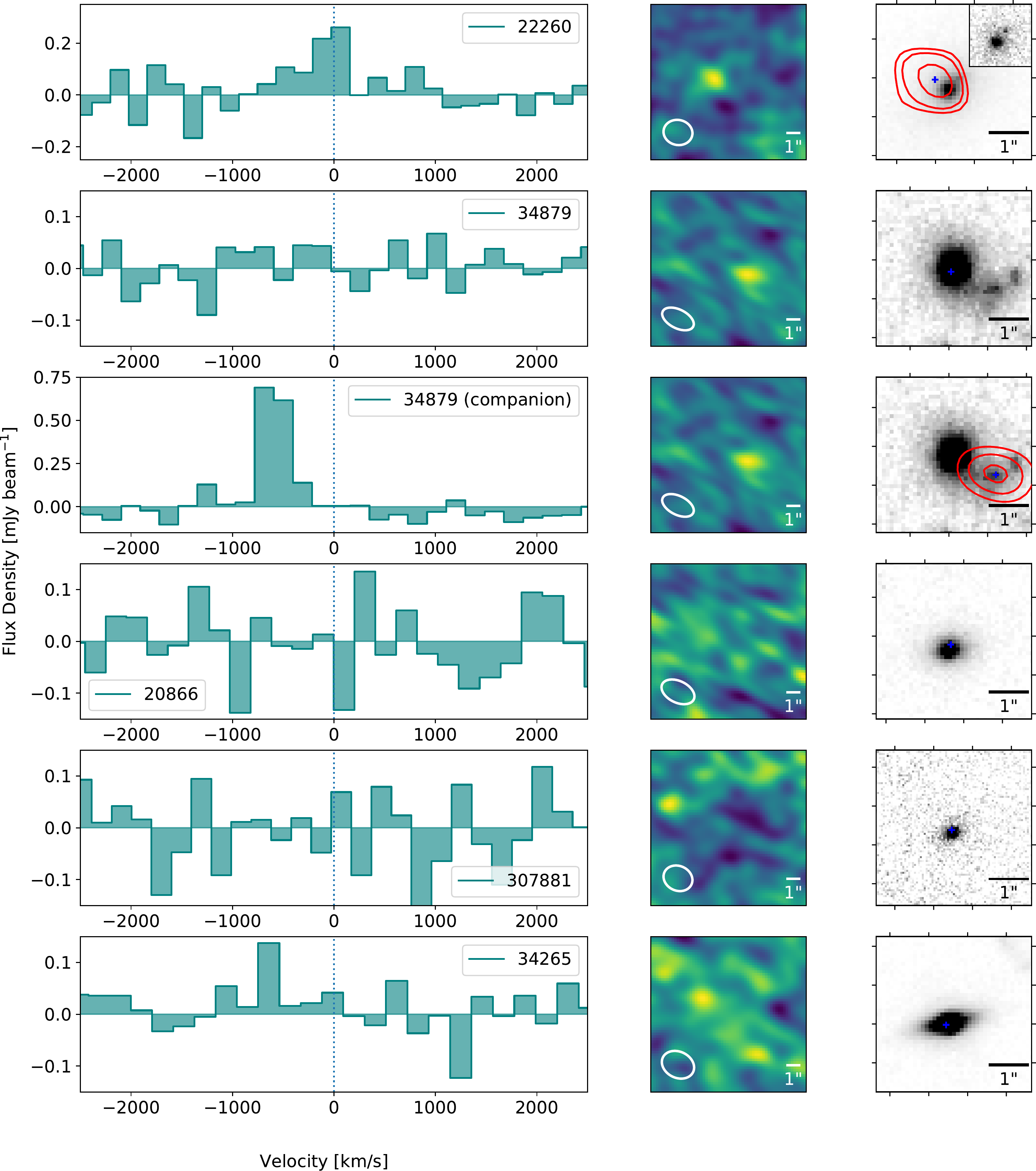}
\caption{ Left panel: ALMA CO(2--1) spectra in 200 km/s channels for each of our galaxies. Spectra are extracted from the position of the  blue cross in right panel. Middle panel: The ALMA CO(2--1) integrated image in 400 km/s channels centered at CO(2--1) of the target galaxy (except 22260 which shows the integrated image in a 500 km/s channel, where we find a 4$\sigma$ detection; we assume the flux 
as originating in our source). ALMA beam is indicated by white ellipse. Right panel: the {\it HST}/WFC3 F160W image for each of our targets. For 22260 we show a zoomed inset of the ACS/F814W imaging where a secondary stellar component is more visible than F160W. We show the CO(2--1) contours in red (where detected). For 22260 we show 50, 60 and 80 mJy/beam km/s contours. For 34879 we show 100, 120 and 130 mJy/beam km/s contours in a 200 km/s channel, offset in velocity from our target galaxy by dv$=$-600 km/s to show the emission of the companion galaxy. 34879 itself is not detected in CO(2--1). 
}\label{fig:spectra}
\end{figure*}

\subsection{Molecular gas measurements}\label{sec:molgas}
To extract CO(2--1) spectra for each source, we used the \texttt{uvmultifit} package \citep{MartiVidal2014} to fit pointlike sources to the visibilities, averaging together channels in order to produce a number of resulting spectra with channel widths ranging from 50--800\,km\,s$^{-1}$. Given the low spatial resolution of the data and the compact galaxy sizes as measured in the available \textit{HST} imaging, the point-like source approximation is likely valid. For most sources, we fixed the position of the point source component to the phase center of the ALMA data, with two exceptions detailed below, leaving only the flux density at each channel as a free parameter. The spectra of each target are shown in Figure~\ref{fig:spectra}.

In source 22260, we detect a weak emission line at the correct frequency for the galaxy's redshift, but offset $\sim$1.2$\pm$0.3'' from the expected position of the target galaxy, a marginally-significant offset given the signal-to-noise of these 2'' resolution data. It is not clear if this offset is spurious, due to an astrometric offset with respect to the \textit{HST} imaging (although unlikely given our careful registration to GAIA), or reflective of a more complex physical scenario with a gas-rich region within this galaxy or a very nearby secondary galaxy as has been seen in high-redshift quiescent galaxies \citep{Schreiber2018JH}. 
For this source, we fit two point sources to the visibilities, fixing the position of one to the phase center (where we find no detection) and the other to the position of the slightly offset source, which is shown in Figure~\ref{fig:spectra}.
We subsequently treat this as a real detection of CO(2--1) from our target. As can be seen in the {\it HST} imaging shown in Figure \ref{fig:spectra} this galaxy has a secondary optical/near-IR component (seen most prominently at {\it HST}/ACS 814W shown as inset),
possibly indicating a recent minor merger. 
Deeper high-resolution ALMA data would be necessary to conclusively determine if the origin of the CO(2--1) emission is the secondary optical-IR component.

\begin{deluxetable*}{lcccccc}[!th]
\tablecaption{Molecular gas properties } 
\tablecolumns{8}
\tablewidth{0pt}
\tablehead{
\colhead{ID} &
\colhead{S$_{\nu}^{a,b}$ } &
\colhead{S$_{\nu}$dv$^b$ } & \colhead{L'CO(2--1)$^b$ } &
\colhead{$\Mgas$$^{c}$} & \colhead{$\fgas$$^{c}$} \\
\colhead{} & \colhead{$\mu$Jy} & \colhead{mJy kms$^{-1}$} & \colhead{10$^{8}$ K km s$^{-1}$ pc$^2$} & \colhead{10$^{9}$ Msun} & \colhead{\%}
}

\startdata
22260$^d$ & 180  $\pm$ 38 &  90$\pm$ 19 & 19 $\pm$ 4 & 10.5 $\pm$ 2.2 & 3.2 $\pm$  0.7 \\
20866 & 47.4 & 23.7 & 7.5 & 12.3 & $<$ 4.3  \\
34879$^d$ & 27.5 & 13.8 & 3.3 & 5.5 & $<$ 2.6 \\
34265$^d$ & 35.1 & 17.6 & 5.9 & 9.8 & $<$ 3.0  \\
21434 & 69.6 & 34.8 & 8.0 & 13.7 & $<$ 5.5  \\
307881 & 37.8 & 18.9 & 5.3 & 8.8 & $<$ 2.1  \\
\hline
Stack & - & 10.3$^e$ & 2.89 & 4.7 & $<$1.6 \\
\enddata

\tablenotetext{a}{Line flux is measured in a 500km/s channel.  }
\tablenotetext{b}{1$\sigma$ upper limits  }
\tablenotetext{c}{3$\sigma$ upper limits, and assuming r$_{21}$ = 0.8 in temperature units, alpha$_{CO}$ = 4.4.   Molecular gas masses can be rescaled under different assumptions as \Mgas$\times$(0.8/r$_{21}$)(\alphaco/4.4) }
\tablenotetext{d}{OII$\lambda$3727 detected in emission with Keck.}
\tablenotetext{e}{Assumes 400 km/s bin. Can be scaled to width 500 km/s by multiplying by $\sqrt{500/400}$.} 
\label{tab:mol}
\end{deluxetable*}

34879 is a similar scenario, although in that case the line emitter is brighter, offset in velocity from the redshift of our target ($\Delta v\sim$600 km/s), and clearly identifiable with a nearby galaxy in the \textit{HST} imaging (Fig.~\ref{fig:spectra}). We again extract spectra by fitting multiple point sources to the visibility data for this field, fixing the positions of the sources to the phase center and the observed position of the line emitter, respectively. After this procedure, the spectra extracted at the phase centers of each field show no evidence of CO emission, although we note that the channel fluxes in these spectra are now slightly correlated with the spectra of the offset sources due to the small sky separations compared to the synthesized beam sizes.

For the detected source, 22260, we measure the line flux by fitting a simple Gaussian to the CO(2--1) spectrum. For galaxies that are not detected in CO(2--1), we set upper limits to the line flux. The undetected galaxies have velocity dispersions, $\sigma$, measured from the rest-frame optical stellar absorption features in the range $\sigma\sim200-370$ km s$^{-1}$ \citep{Bezanson2013, Belli2014a} with the exception of 307881 for which it was not measured \citep{Onodera2012}. To measure upper limits to the integrated CO(2--1) line flux of undetected galaxies, we assume similar line widths for the stars as any molecular gas, and adopt typical values for the FWHM of the CO(2--1) of 2.355$\times\sigma\sim$500-600 km s$^{-1}$.  We use these line widths and the channel noise to set upper limits on the integrated line fluxes of each target. We note that large linewidths are conservative, and that these upper limits scale with velocity interval $\Delta v$ as $\sqrt{\Delta v}$. Assuming a smaller linewidths would decrease our limiting integrated flux. The CO(2--1) line luminosity for our detected galaxy 22260, and the 1$\sigma$ upper limits in the case of non-detections, are reported in Table \ref{tab:mol}.

To convert our measurements of CO(2--1) line luminosities into molecular gas mass (\Mgas), we make the following standard assumptions about the molecular gas conditions. 
We first assume a CO excitation (namely the luminosity ratio between the CO(2--1) and CO(1-0) transitions, r$_{21}$) following observations from the local Universe. Although local galaxies with low sSFR are observed to exhibit a range of values r$_{21}=0.7-1$ \citep[e.g.][]{Saintonge2017}, bulges and the central nuclei of galaxies that are thought to be similar to high-redshift compact quiescent galaxies exhibit near-thermalized excitation, r$_{21}=1$ \citep[e.g.][]{Leroy2009}. For our analysis we assume r$_{21}=0.8$ following \citet{Spilker2018}, which results in more conservative (higher) molecular gas mass measurements and limits than the assumption of thermalized emission. For comparisons to other measurements in the literature of $z>1$ passive galaxies we therefore rescale other values to this excitation as \Mgas$\times$(0.8/r21) (including the object from \cite{Bezanson2019} which we convert to our value of r$_{21}=0.8$). Assuming a larger value for r$_{21}$ (e.g. 1, as is done for other studies of passive systems across redshifts) does not significantly change our results, and instead would imply even lower molecular gas fractions that further strengthen our conclusions. 
This assumption has a minimal impact on our \Mgas\ uncertainty budget (10-20\%) compared to e.g. a factor of $\gtrsim$2 uncertainty due to the assumed value of the CO-H$_{2}$ conversion factor to translate the measured CO luminosity to \Mgas.

In this work we assume a Milky Way like value of \alphaco = 4.4 \msol (K km s$^{-1}$pc$^{2}$)$^{-1}$, which is a reasonable assumption for massive galaxies with presumably high metallicities  (e.g. \citealt{Narayanan2012}; see also the review by \citealt{Bolatto2013}).

\subsection{Stack of non-detections in CO(2--1)}

With five out of six galaxies undetected in CO(2--1) \citep[including 21434;][]{Bezanson2019}, we perform a stacking analysis of the five non-detected galaxies. We calculate the weighted average (mean) to account for the slight differences in map rms, and use the non-primary beam corrected maps, which have Gaussian noise properties.
Since the nearby companion of galaxy 34879 has significantly detected CO(2--1) emission offset by 600 km/s, but with roughly width of 200 km/s, we restrict our exploration of stacked CO(2--1) emission using image cubes with velocity resolution $\lesssim$400 km/s to prevent the flux from the companion entering the stack, given the companion's location within 1.5'' of the target galaxies in the stack. We construct image cubes at 400km/s velocity resolution centered at the rest-frequency of CO(2--1) of each galaxy, and stack the velocity bin that contains the CO(2--1) line.

We do not detect any CO(2--1) from the stack of individually undetected sources, with an rms noise limit of 25.6$\mu$Jy/beam for the 400\,km/s channel width of the stack. We use the mean redshift of the non-detected galaxies ($<z>$=1.476) to put a 1$\sigma$ upper limit to the average CO luminosity of \lprime$_{(2-1)}<2.9\times10^{8}$ K km s$^{-1}$ pc$^{2}$. We make the same assumptions listed in Section \ref{sec:molgas} to convert this measurement to a molecular gas mass and find \Mgas$<4.7\times10^{9}$\msol (3$\sigma$). Using the average stellar mass of our undetected sample of log$_{10}\Mstar/\msol\sim11.5$, this puts a 3$\sigma$ upper limit on the molecular gas fraction of 1.6\%. The stacked sample has an average specific star formation rate of $6\times10^{-11}$ yr$^{-1}$ (likely an upper limit, as explained in Section \ref{sec:sfr}) and the properties derived from the stack are summarized in Table \ref{tab:mol}.

\section{Results}\label{sec:results}
Our new ALMA observations indicate that our sample of massive (log$_{10}\Mstar/\msol > 11.3$) and quiescent (log$_{10}$ sSFR$\lesssim -10$ yr$^{-1}$)  galaxies at $z>1$ have low molecular gas masses (\Mgas$\lesssim5-10\times10^{9}$\msol),  translating to molecular gas fractions (\fgas = \Mgas/\Mstar) between $\sim$2-6\%. To provide context for these measurements, we compile measurements of molecular gas using CO as a tracer from the literature across redshifts. We include surveys that target low-J transitions (J$_{up}\lesssim2$) to minimize uncertainties from variations in the methods to correct for CO excitation. The majority of these surveys targeted star forming galaxies outside the local Universe, including PHIBSS \citep{Tacconi2013}, PHIBSS2 \citep{Tacconi2018, Freundlich2019}, as well as smaller programs targeting Milky-Way progenitors \citep{Papovich2016}, extended disk galaxies \citep{Daddi2010}, compact star-forming galaxies \citep{Spilker2016}, and galaxies from overdense regions \citep{Hayashi2018,Rudnick2017}. 
To targeted samples, we add CO-detected sources from the blind ASPECS Survey \citep[][]{Decarli2016, Aravena2019}. 
We also include the few studies that have targeted quiescent or post-starburst galaxies outside the local universe at $z<1$ \citep{Suess2017,Spilker2018}. Finally we include the large surveys at $z\sim0$ that have enabled an exploration of molecular gas in similarly massive galaxies to our sample, at similarly low sSFRs \citep[albeit at late cosmic times;][]{Young2011,Saintonge2012,Saintonge2017,Davis2016}.

To date, molecular gas measurements using CO exist for only two confirmed quiescent galaxies above $z>1$; these are upper limits (3$\sigma$) on a massive quiescent galaxy published by  
\citet[][$\fgas\lesssim13$\%, converted from Salpeter IMF]{Sargent2015} 
and the pilot galaxy for this survey 
($\fgas\lesssim5.5\%$; \citealt[][ using our derived stellar mass and velocity width, to be consistent with the rest of the sample]{Bezanson2019}). Both measurements are rescaled to our assumption r$_{21}$=0.8. Both galaxies are spectroscopically confirmed, enabling a robust upper limit to their molecular gas content.

\begin{figure}[th]
\includegraphics[scale=0.85, trim=15 9 3 10,clip]{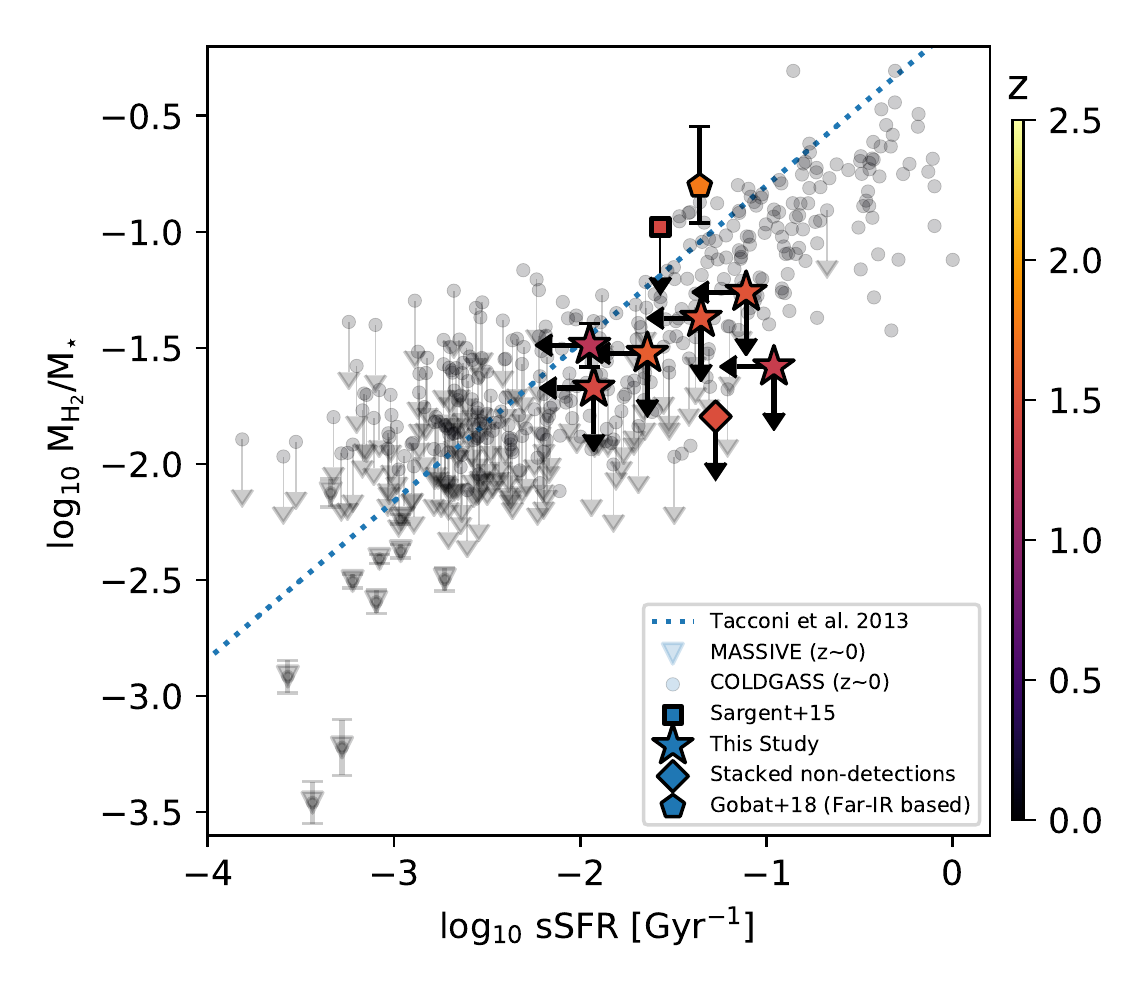}
\caption{ Comparison of our CO measurements to those of quiescent galaxies at $z\sim0$ from the COLDGASS and MASSIVE surveys \citep{Saintonge2017, Davis2016}.
Large symbols indicate quiescent galaxies at $z>1$ \citep[this work;][]{Sargent2015, Bezanson2019} and the far-infrared based stack of \citet{Gobat2018}.
Our sample has low \fgas $<2-6 \%$; comparably low to galaxies at z=0 with similarly low sSFR.  }\label{fig:fgas_vs_z0}
\end{figure}

The most comprehensive constraint on the average molecular gas in quiescent galaxies at $z>1$ to date is a far-infrared stack of 977 photometrically selected quiescent galaxies \citep{Gobat2018}, where the molecular gas content is inferred from the average dust emission \citep{Magdis2012}. 
We add this measurement to the CO constraints from the literature because it uses the largest sample of quiescent galaxies at $z>1$.

\begin{figure*}[]
\includegraphics[scale=0.85, trim=10 9 35 10,clip]{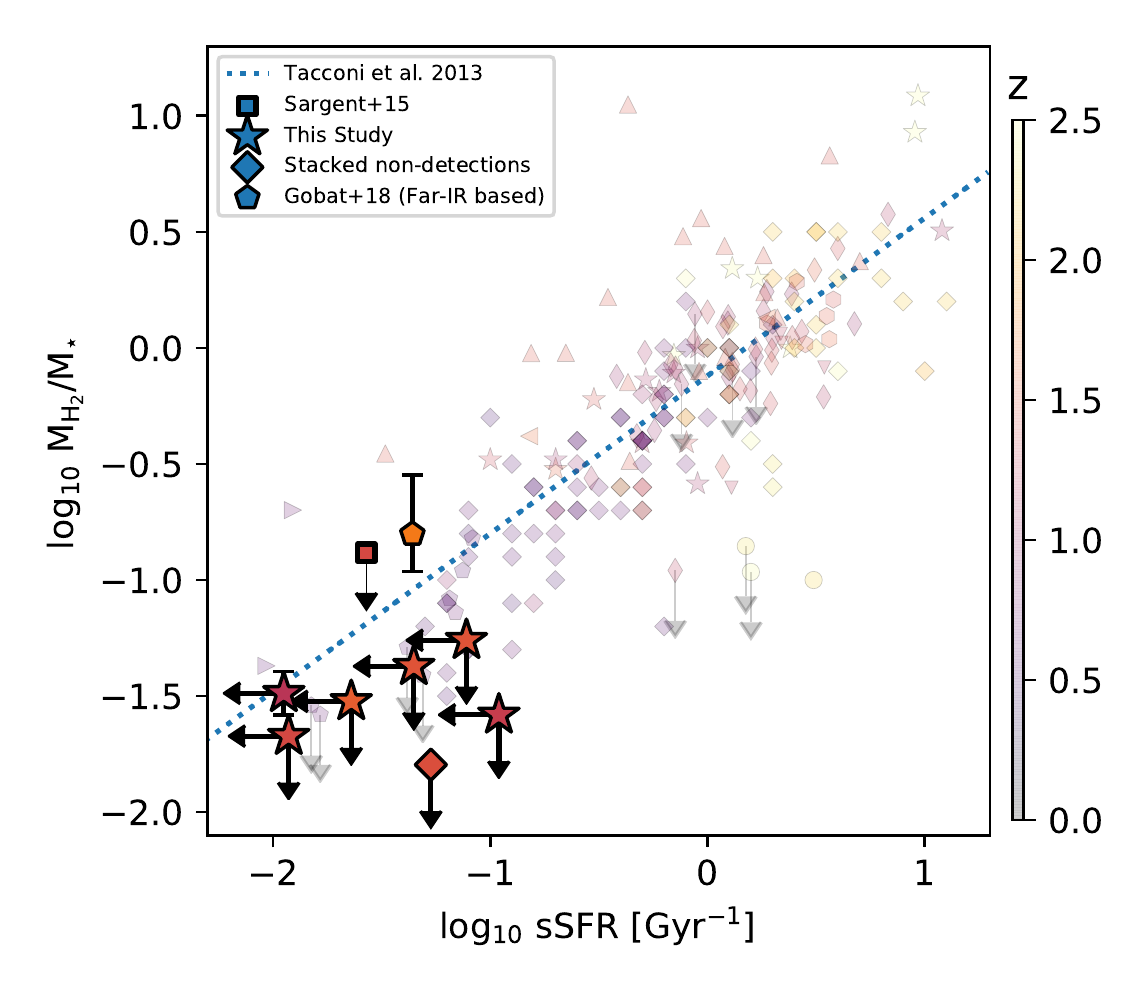}
\includegraphics[scale=0.85, trim=10 9 3 10,clip]{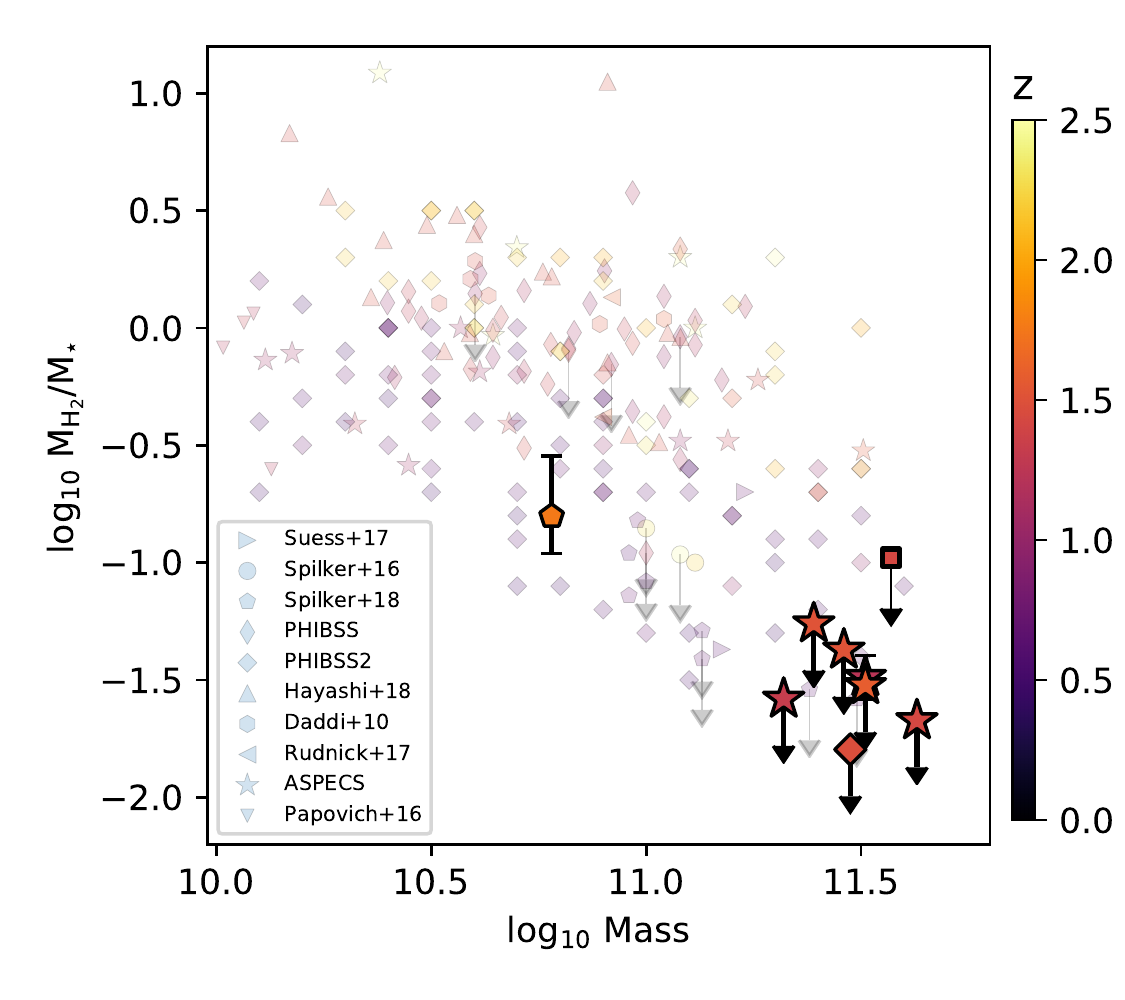}
\caption{
Comparison of our measurements to measurements based on CO in literature at $z>0.5$.
Large symbols indicate quiescent galaxies at $z>1$ \citep[this work;][]{Sargent2015, Bezanson2019} and the far-infrared based stack of \citet{Gobat2018}.
Small symbols (defined in right panel legend) indicate comparison literature measurements. 
Our sample have low molecular gas fraction $<2-6 \%$; 1-2 orders of magnitude lower than few coeval star-forming galaxies at similar stellar mass.  }\label{fig:fgas_vs_m}
\end{figure*}

Figures~\ref{fig:fgas_vs_z0} and \ref{fig:fgas_vs_m} show \fgas\ for our sample as a function of sSFR and \Mstar. We plot our 
ALMA measurements as stars, along with the measurements from the literature (small translucent symbols). Quiescent galaxies at $z>1$ are large bold symbols. We additionally include the stacked measurement from the five non-detected galaxies (diamond). Figure \ref{fig:fgas_vs_z0} shows that based on our deep limits, massive and quiescent galaxies at $z>1$ have comparably low gas fractions relative to galaxies at $z=0$ with similar sSFRs, and that our deep \fgas\ limits are comparable to local surveys.
Figure \ref{fig:fgas_vs_m} shows that our upper limits on \fgas\ are the lowest CO-derived constraints on molecular gas content of any galaxy population above $z>1$.

The left panel of Figure \ref{fig:fgas_vs_m} shows \fgas\ vs sSFR at $z>0$, with galaxies color coded by redshift. 
The gas fraction measurement/limits for our sample are about an order of magnitude deeper than the limit set by \citet[][]{Sargent2015}, and an order of magnitude lower than that inferred from dust emission by \citet[][]{Gobat2018}. We discuss this discrepancy in quiescent galaxy \fgas\ between their average detection and our deep limits further in Section \ref{sec:scatter}.

In the right panel, we plot \fgas\ vs stellar mass, where galaxies are again color coded by redshift. Our measurements are in line with observations that the gas fraction in galaxies decreases with increasing stellar mass at all redshifts, although the mass dependence is weak compared to the stronger dependencies on redshift and sSFR \citep[e.g.][]{Tacconi2018}. Our study doubles the number of constraints on molecular gas mass at z $>$1 at the  massive (log$_{10}$\Mstar/\msol $>$ 11.3) end.

In Figure \ref{fig:fgas_vs_z} we plot \fgas\ as a function of redshift, where galaxies are color coded by sSFR. A number of well-known scaling relations are apparent in Figures \ref{fig:fgas_vs_z0}, \ref{fig:fgas_vs_m}, and \ref{fig:fgas_vs_z} including that overall, the molecular gas fractions in galaxies decrease with decreasing redshift, decreasing sSFR, and increasing stellar mass. Our data contributes new datapoints to the poorly explored parameter space at low sSFR, high mass, at high redshift.

\section{Discussion}\label{sec:discussion}

In this paper, we have placed constraints on the molecular gas content in the first sample of massive quiescent galaxies at $z>1$ ($<z>=1.45$).
Our low \fgas\ measurements indicate that the exhaustion or destruction of molecular gas in massive quiescent galaxies is efficient and complete, consistent with the finding for our pilot galaxy \citep{Bezanson2019}. 
That massive quiescent galaxies at $z>1$ are gas poor suggests high star-formation efficiency and rapid depletion times during their evolution.
While our sample is not complete in stellar mass, we do not find evidence within our sample that \fgas\ varies with either galaxy size or surface density $\Sigma_*$. Among quiescent galaxies, these structural properties are known to correlate with stellar age \citep[e.g.][]{Williams2017}, formation redshift \citep[e.g.][]{EstradaCarpenter2020} and quenching timescale \citep[e.g.][]{Belli2019}, and therefore plausibly trace timescales for gas consumption.  We measure \fgas$<2-6\%$, values that are universally low despite the large dynamic range of structure we probe among quiescent galaxies at $z>1$ ($R_e=0.9-7$ kpc; log$_{10}\Sigma_* = 8.9-10.8$ \msol kpc$^{-2}$; Figure \ref{fig:props}).\footnote{We note that quiescent galaxies are generally higher $\Sigma_*$ than star forming galaxies, which have larger gas reservoirs.}

Our sample suggests that massive galaxies that cease star formation at the peak epoch of quenching do not retain large reservoirs of gas. These findings are in contrast with observations of recently quenched galaxies at $z<1$, some of which contain significant molecular gas reservoirs (\fgas$\sim20-30$\%), suggesting that their low SFRs are due to decreased star-formation efficiency (e.g. suppressed dynamically) rather than a lack of fuel for star-formation \citep{Rowlands2015, French2015, Suess2017, Smercina2018, Li2019}.  Furthermore,  \citet[][]{Spilker2018}  find \fgas$<~1-15$\% in quiescent galaxies at intermediate redshifts ($z\sim0.7$), additional evidence for heterogeneity among galaxies below the main sequence. Our new results, collectively with those at $z<1$, highlight a diversity in molecular gas properties among quenching galaxies across cosmic time, possibly indicating that the primary drivers of quenching change over cosmic time. These new observations of the variation in gas reservoirs of non-starforming galaxies across cosmic time are therefore important constraints for our theoretical formulations of quenching processes, and the time evolution of gas reservoirs. In the following sections, we explore the implications of our new low gas fraction measurements in this context.

\subsection{The distribution (intrinsic scatter) of cold gas content among $z>1$ quiescent galaxies}\label{sec:scatter}

Although this is the first systematic study using CO to measure molecular gas  in quiescent galaxies at $z>1$, 
the recent observation of average far-IR properties of 977 quiescent galaxies at $z>1$ found significant dust continuum emission, implying a relatively large molecular gas content \citep[\fgas$\sim16$\% when converted to Chabrier IMF;][]{Gobat2018}. The individual measurements of molecular gas in our quiescent sample range from \fgas$\lesssim2-6$\%, and are inconsistent with the average \fgas\ measurement by \citet[][]{Gobat2018}. While a primary uncertainty in our \fgas\ measurement is the assumed value of \alphaco, extreme values only observed in low mass and low metallicity systems  \citep[\alphaco$\gtrsim15$;][]{Bolatto2013, Narayanan2012} would be required to bring our measurements into agreement.

A direct comparison to the Gobat et al. result is difficult owing in part to our differing methodologies, each of which is subject to its own systematic uncertainties.  
And, as with any photometric selection of quiescent galaxies, there is always some risk of contamination from dusty star-forming galaxies. The contamination may enter the stack either through misidentification because of the age-dust degeneracy of colors (even if only a few bright objects), or due to neighboring dusty galaxies given the low spatial resolution of the far-IR data ($\sim$10-30''). Neighbors can contaminate either through poor source subtraction, or as hidden dusty galaxies that do not appear in optical/near-IR selection but may remain within the far-IR photometric beam \citep[e.g.][]{Simpson2017, Schreiber2018JH, Williams2019}. Further, both our sample and dusty star forming galaxies are massive and may be strongly clustered \citep[e.g.][although see also \citealt{Williams2011}]{Hickox2012}.
In this section we ignore any such possible contamination, and discuss several physical explanations for this disagreement.

First, the relatively large \fgas\ observed by \citet[][]{Gobat2018} could be reflecting a heterogeneity of molecular gas properties among the passive galaxy population at $z>1$, as is observed at $z<1$. Our sample represent some of the most massive and oldest passive galaxies known at $1<z<1.5$, while the \citealt{Gobat2018} sample is dominated by objects less massive than our sample ($<$log$_{10}$\Mstar$> =$ 10.8). Perhaps lower mass and/or younger additions to the red sequence still have molecular gas leftover, contributing to the far-IR emission observed on average. 
However, we note that because the stack is average, and our measurements are $>10\times$ lower \fgas, a heterogenous sample would imply even larger $\fgas>16\%$ in any sample of gas-rich quiescent galaxies. 
More surveys that span a larger range of parameter space for individual quiescent galaxies (e.g. lower mass) are required to investigate this explanation further (Whitaker et al. in prep, Caliendo et al. in prep).

Alternatively, the calibration to convert the far-IR emission into a measurement of \Mgas\ might not be universal.
These conversions are typically based on assumed dust to gas ratios and/or dust temperatures, calibrated using primarily star-forming galaxies \citep[e.g.][]{Magdis2012, Scoville2016}.  In theory this relies on an intrinsic relationship between dust and gas content that has been shown to accurately describe star forming galaxies \citep{Kaasinen2019}, and for the most part, also holds for quiescent galaxies in the local Universe, albeit with large scatter \citep[e.g.][]{Lianou2016}. In principle, dust traces both atomic (HI) and molecular (H$_{2}$) gas phases, and so this could still hold if the HI/H$_{2}$ ratio is high in quiescent galaxies, while the dust to H$_{2}$ ratio is very low. We note that \citet{Spilker2018} stacked the 2mm dust continuum emission to compare to \Mgas\ measured from CO, finding consistent values between \Mgas\ observed via CO and dust, lending support for the idea that dust to H$_{2}$ conversions hold for massive quiescent galaxies at high redshift. However, \citet{Gobat2018} make the simplifying assumption that all gas traced by dust is molecular, although HI/H$_{2}$ mass ratios in local quiescent galaxies can be large \citep{Zhang2019} and diverse \citep{Welch2010,Young2014,Boselli2014, Calette2018}. It is therefore a possibility that the significant dust emission detected by \citet[][]{Gobat2018} is not in conflict with our low \fgas\ measurements, and instead is primarily tracing atomic HI rather than H$_{2}$.

Nevertheless, other factors may affect these conversions, warranting further exploration. For example the dust to gas ratio can also vary with metallicity, as explored in simulations, although the extent to which this disrupts scaling relations is not clear 
\citep[i.e. gas/dust may plateau above solar metallicity, applicable to most massive galaxies;][]{Privon2018, Li2019}.  Future samples of quiescent galaxies with observations of both CO and dust continuum emission would reveal if the dust to \Mgas\ calibrations apply across galaxy populations at high redshift, as done locally \citep{Smith2012}.
The comparison between our work presented here and that presented in \citep{Gobat2018} thus highlights several avenues of future investigation to understand the intrinsic scatter in molecular gas properties of quiescent galaxies, which will help understand the diversity of pathways that passive galaxies may take to quiescence. 

\subsection{Timescales for gas consumption or destruction}\label{sec:timescale}
Accretion is now considered to be a primary driver of galaxy growth in the early Universe \citep[for a review see][]{Tacconi2020}. While observations support 
this picture, it remains unclear what disrupts the growth in massive galaxies that become quiescent. Explanations include the destruction or expulsion of gas due to feedback, the suppression of gas accretion (e.g. by virial shocks once log$_{10}$\Mhalo/\msol$>$12), or the suppression of gas collapse due to the development of a stellar bulge. Our observations of molecular gas in quenched galaxies can help discriminate between the different processes.  
In particular, we explore here the timescales for gas expulsion or consumption that are consistent with the low gas fractions we measure. Unfortunately with mostly upper limits to \Mgas, and likely only upper limits to the SFR, our dataset precludes a robust measurement of (current) depletion times (\tdep = \Mgas / SFR) and we instead explore the allowable range of \tdep\  given low \fgas\ and old mass-weighted stellar age.

\begin{figure*}[]
\begin{center}
\includegraphics[scale=0.88, trim=10 9 70 10,clip]{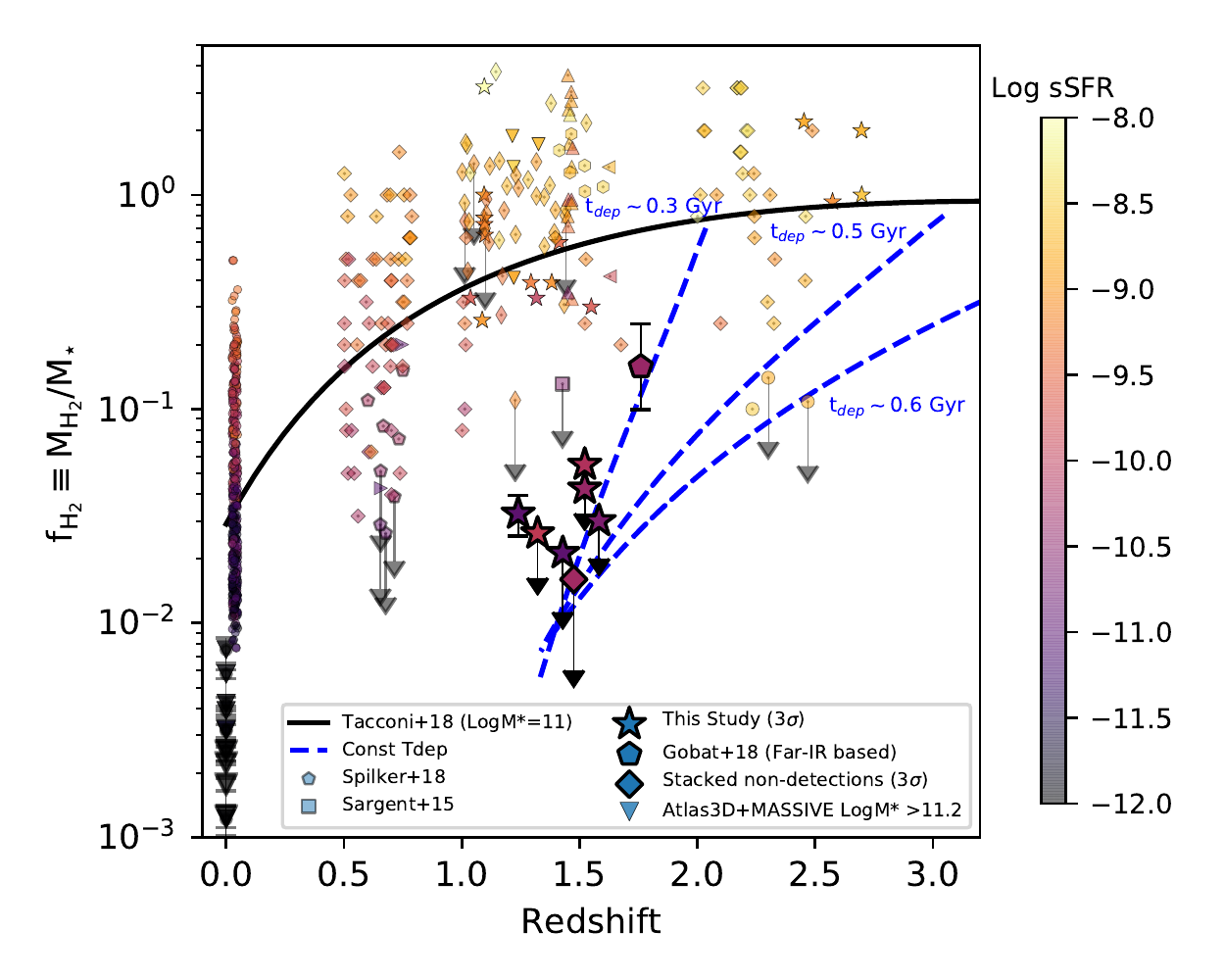}
\includegraphics[scale=0.88, trim=55 9 10 10,clip]{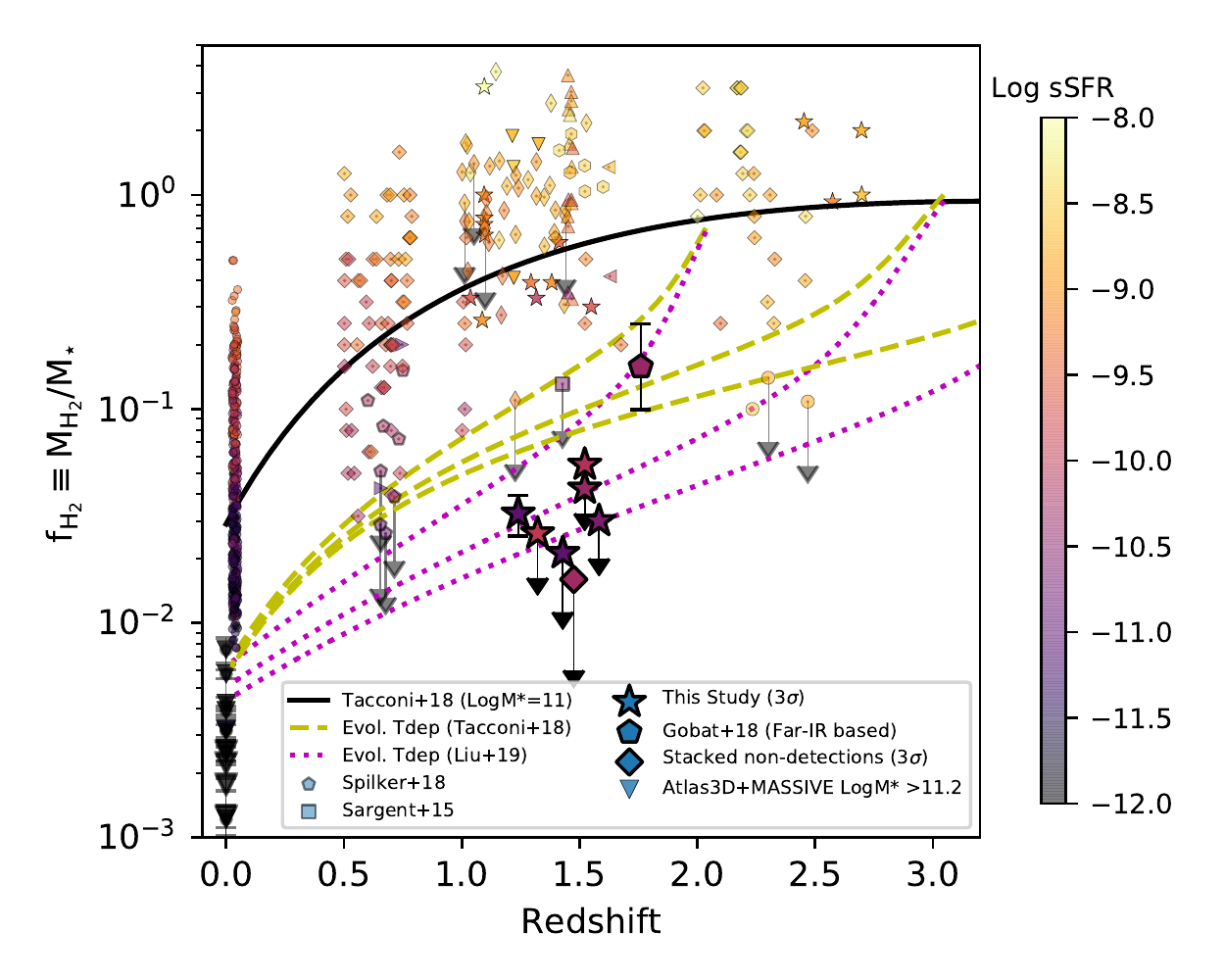}
\caption{\fgas vs redshift for galaxies in our sample (large stars) and literature measurements. All galaxies are color-coded by their sSFR. Symbols are represented as in Figure \ref{fig:fgas_vs_m}. For clarity we omit $z=0$ measurements below \fgas $<$1e-3 from the Atlas3D or MASSIVE surveys, and two measurements above \fgas$>$5 at $z\sim2$ from \citep{Hayashi2018}. Black line indicates the \fgas\ on the main sequence for star forming galaxies with log$_{10}$\Mstar/\msol$=11$. 
Lines show the gas depletion according to our toy models outlined in Section \ref{sec:timescale}: blue curves indicate models with constant \tdep=0.3, 0.5, 0.6 Gyr where accretion halts at $z=2,3,4$, respectively. Yellow indicate models where the value of \tdep\ changes according to scaling relations measured by \citet[][]{Tacconi2018}. 
Our low gas fractions require rapid \tdep, inconsistent with \citet{Tacconi2018}, and have better agreement with relations that have faster depletion times at high redshift, low sSFR, and high mass \citep[][magenta curves]{Liu2019}. 
}\label{fig:fgas_vs_z}
\end{center}
\end{figure*}

\subsubsection{Closed-box toy model: constant \tdep}
To provide qualitative insight into 
the timescales required to achieve the low \fgas\ we observe, 
we construct a closed-box toy model for a log$_{10}\Mstar/\msol\sim11$ main sequence galaxy that stops gas accretion, and then depletes its existing gas reservoir at specified depletion times. The SFR is decreased accordingly as gas is consumed. This model is qualitatively similar to that used in \citet[][]{Spilker2018} to investigate if their measured depletion times for quiescent galaxies at $z\sim$0.7 are consistent with depleting to levels observed in quiescent galaxies at $z=0$.

We first assume a toy model with a constant \tdep\ that remains the same with time and SFR, and calculate how the gas fraction declines if the gas accretion is halted while the galaxy is on the main sequence at $z=2,3,4$.  We additionally assume that as the galaxy consumes its gas through star formation, stellar mass loss will return $\sim30\%$ of that mass back to the interstellar medium \citep[ISM; for a Chabrier IMF; e.g.][]{LeitnerKravtsov2011, Scoville2017}. While this is a physically motivated assumption, it also is conservative. If the true fraction of gas returned to the ISM is lower, the gas reservoir will be depleted even faster, strengthening our conclusions.

In the left panel of Figure \ref{fig:fgas_vs_z}, the blue curves show the evolution of these constant \tdep\ closed box models at $z=2,3,4$ for \tdep$=0.3, 0.5, 0.6$ Gyr, respectively.
The higher the redshift that accretion is halted, the longer the limiting \tdep\ that is consistent with our low gas fractions. Longer depletion times will flatten the blue curves and are inconsistent with our measurements. These curves indicate that rapid depletion times are required for a main sequence galaxy to use up its existing reservoir if accretion is halted. The earlier in cosmic time the accretion is halted, the longer \tdep\ can be and still match our observations. However we note that for mass-weighted ages of $\sim$1-3 Gyr (all galaxies except 22260\footnote{We note the possibility that 22260  received its gas later through a minor merger from its secondary component, and therefore its older age does not disagree with this picture.}), the majority of star-formation occurred before z$=3.5$, indicating a limiting \tdep$<0.6$ Gyr. This is also roughly the typical depletion times for massive galaxies on the main sequence at these redshifts  \citep[$\sim$0.4-0.6 Gyr][]{Tacconi2018,Liu2019}. 
Our data is consistent with this simple picture where galaxies truncate accretion and then consume the existing gas at typical main sequence \tdep\ rates, or faster.

\subsubsection{Closed-box toy model: varying \tdep
}

While the constant \tdep\ toy model is useful for providing the qualitative intuition that long depletion timescales (\tdep$ > 0.6$ Gyr) are inconsistent with our data, observations have shown that in reality, \tdep\ is not constant as galaxies evolve. \tdep\ is known to vary as a function of redshift, sSFR (i.e. distance from the main sequence), with weaker dependences on \Mstar\ and galaxy size \citep[][]{Tacconi2013, Santini2014, Genzel2015, Tacconi2018, Liu2019, Tacconi2020}. Therefore, we also explore a closed box model where the \tdep\ smoothly evolves according to scaling relations as galaxies leave the main sequence. These scaling relations imply an increase in \tdep\ as galaxies move below the main sequence, which slows the rate that \fgas\ decreases with time. For this set of toy models we make the conservative, albeit unrealistic, assumption that no mass is returned by stars to the ISM as galaxies move below the main sequence. This is the more conservative comparison in this case, because any mass loss to the ISM during this phase will increase the time required for the toy model to reach the low \fgas\ we observe. 

The right panel of Figure \ref{fig:fgas_vs_z} shows the result of this toy model calculation for two example scaling relations, that of \citet[][]{Tacconi2018} in yellow and of \citet[][]{Liu2019} in magenta. 
In the case of \citet{Tacconi2018}, the simple consumption of gas does not reach low enough gas fractions quickly enough to match our observations. This is due to the relatively long depletion times below the main sequence implied by this particular scaling relation. This is not necessarily surprising, as primarily star-forming galaxies are used to calibrate these relations outside the local Universe; understanding the evolution of the star-forming population was the primary goal of these analyses. Scaling relations measured in \cite{Genzel2015,Tacconi2020} result in similar behavior.

\begin{figure*}[]
\begin{center}
\includegraphics[scale=1.3, trim=10 9 10 10,clip]{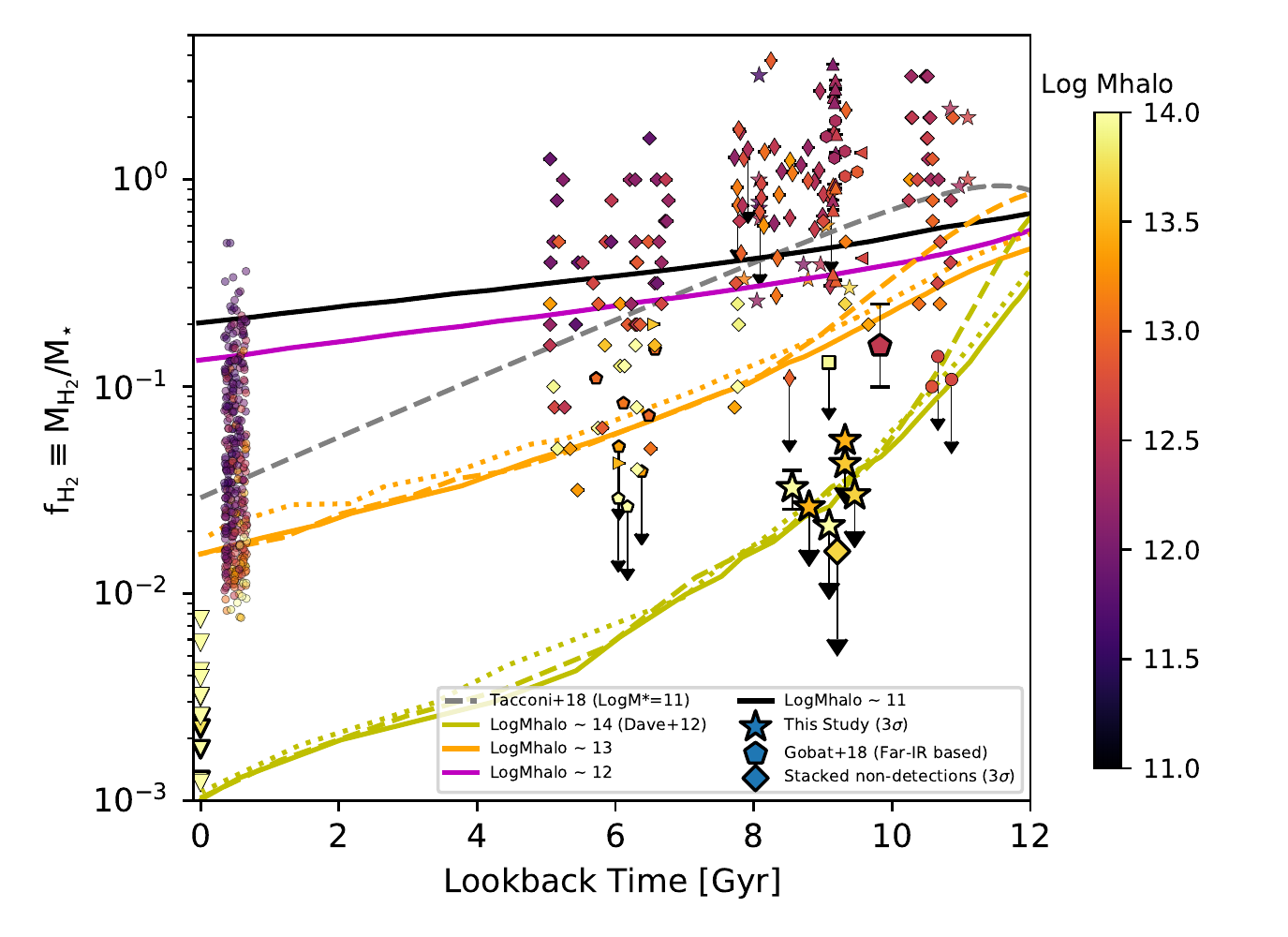}
\caption{ Same as Figure \ref{fig:fgas_vs_z} but with \fgas(z) predictions from analytical equilibrium ``bathtub models" that balance gas inflow, outflow, and star-formation.  
Curves represent halos with masses at $z=0$ of \Mhalo$=$10$^{11}$ (black), 10$^{12}$ (magenta), 10$^{13}$ (orange) and 10$^{14}$ 
(yellow) \msol\ published in \citet[][]{Dave2012}. Solid lines indicate a model where the mass loading factor for outflowing gas is similar to momentum driven feedback. 
For halos with final mass $>$10$^{13}$ we plot two other feedback prescriptions (dotted and dashed lines, see text) but our conclusions are independent of the stellar feedback prescription.
Our data are most consistent with massive halos (\Mhalo=10$^{14}$\msol at z=0) which reached the critical halo mass \Mhalo=10$^{12}$\msol the earliest, and z$\sim4$ (to slow accretion of baryons due to shock heating at the virial radius).
}\label{fig:fgas_dave}
\end{center}
\end{figure*}

Taken at face value, the 
\citet{Tacconi2018} relation implies that \tdep=0.7, 0.6, 0.5  Gyr for a log$_{10}$\Mstar/\msol=11 galaxy leaving the main sequence at z=2,3,4, as explored in Figure \ref{fig:fgas_vs_z}. Extrapolating the relation to the average mass, sSFR and redshifts of our ALMA targets would imply \tdep$\sim$1.6 Gyr and \fgas$\sim$10\%. For the ALMA galaxies individually, the relation implies \fgas\ values that are $>2\times$ larger than our conservatively measured 3$\sigma$ upper limits. Our data safely rule out these extrapolations.

In contrast, the closed box model based on scaling relations measured by \citet[][]{Liu2019} reaches substantially lower \fgas. This is primarily due to a more rapid \tdep\ near but below the main sequence at high redshift in their calibration, compared to the behavior of \tdep\ measured by \citet{Tacconi2018}. As is apparent, the faster \tdep\ near but below the main sequence has a substantial impact on the behavior of our toy model. Therefore, we cannot rule out that our low \fgas\ are consistent with simple gas consumption with behavior similar to \citet[][]{Liu2019} if accretion onto galaxies is halted.  However, we note that including physically motivated gas recycling from stellar mass loss as in the last section would drastically increase the time required to reach our measured \fgas, and increasing the tension with our observations (30\% gas recycling as assumed in the previous section results in the $z=4$ magenta curve consistent with only our two highest \fgas\ constraints, too high to explain all our measurements. The $z=2-3$ curves would be inconsistent with all of our data). 

At face value, the 
\citet{Liu2019} relation implies that \tdep=0.5, 0.4, 0.3  Gyr for a log$_{10}$\Mstar/\msol=11 galaxy leaving the main sequence at z=2,3,4, as explored in Figure \ref{fig:fgas_vs_z}. Extrapolating to the properties of our ALMA targets would imply longer \tdep$\sim$2.6 Gyr and lower \fgas$\sim$6\%, closer to our observations but still $\sim4\times$ larger than our stacked result.

We note that our 3$\sigma$ upper limits and our assumption about r$_{21}$ are conservative, and thus the real gas fractions are likely much lower than the figure suggests. Therefore, we speculate that \tdep\ must remain rapid, in disagreement with extrapolations from scaling relations, as galaxies move below the main sequence. 
Unfortunately, our toy model is highly sensitive to the form of scaling relations at high masses, high redshifts, and below the main sequence, which is poorly explored parameter space. 
This highlights the need for further exploration of gas reservoirs in galaxies below the main sequence in the early Universe.

Our finding that scaling relations do not describe galaxies below the main sequence (at least outside of the local Universe) is in agreement with findings by \citet[][]{Spilker2018}. 
Half of their sample (the half with higher log$_{10}$sSFR $>-1.2$ Gyr and lower mass log$_{10}\Mstar/\msol\lesssim11$) was detected in CO with \fgas$\sim7-15$\%, in agreement with scaling relations.
However, the \fgas\ limits measured in their non-detected sample (with similar sSFR and \Mstar\ to our sample) were significantly lower than the expectations based on scaling relations. Both our data and that of \citet{Spilker2018} indicate that scaling relations for the star-forming population don't extrapolate to populations with lower sSFR, and break down around $3-5$ times below the main sequence \citep{Spilker2018}. Rather, \tdep\ likely remains short below the main sequence until gas is used and destroyed, and what little is left cannot be efficiently converted into stars, thereby increasing \tdep.

Despite the uncertainty in behavior of scaling relations below the main sequence at high-redshift, these comparisons are useful because they qualitatively indicate that \tdep\ must be rapid when galaxies are shutting off their star-formation. These conclusions are the same whether we assume the galaxy originates on the main sequence or in a starburst phase \citep[with even faster typical \tdep; e.g.][]{Silverman2015, Silverman2018}. Smoothly evolving models of departure from the main sequence where star-formation efficiency is decreased and \tdep\ is increased (i.e. reservoirs of gas exist but do not form stars) are inconsistent with our observations. 

Finally, we note the possibility that \tdep\ evolution is not smooth, and an initial rapid drop in gas fraction due to, e.g. increased star formation efficiency or feedback as galaxies go below the main sequence, is followed by an extended period of low \fgas\ and long \tdep. That long depletion times kick in after most gas is gone is also consistent with simulations presented by \citet{Gensior2020} that indicate that suppression of star formation efficiency (i.e. lengthening of \tdep) due to dynamical stabilization by growth of a bulge in galaxies below the main sequence has an impact only at low \fgas \citep[$<$5\%; see also][]{Martig2009, Martig2013}. Such a scenario implies an even faster initial depletion of gas than we model here. Therefore, the \tdep\ values derived for our sample from these toy models should be considered upper limits.

\begin{figure*}[]
\begin{center}
\includegraphics[scale=1., trim=10 0 10 10,clip]{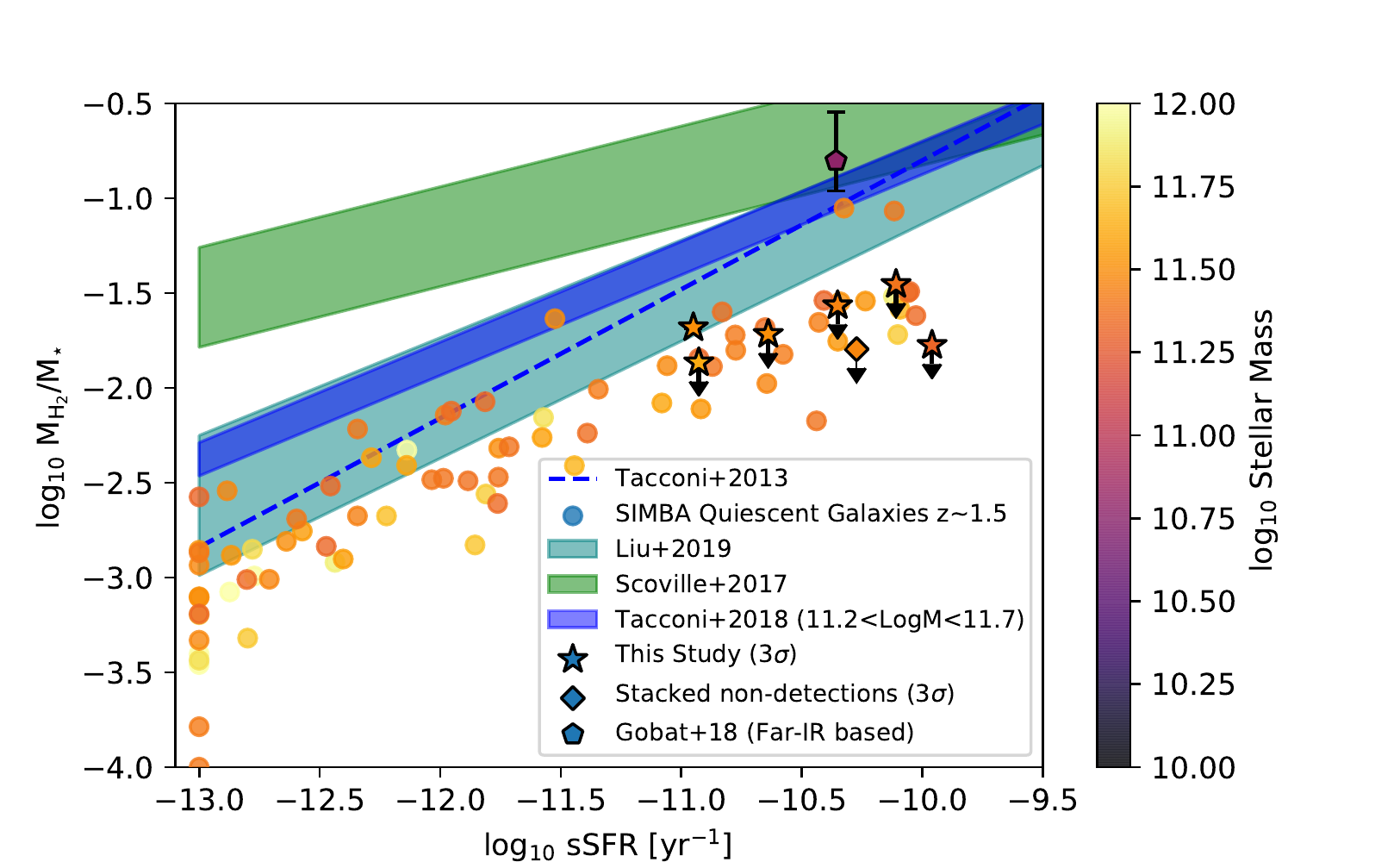}
\label{fig:fgas_simba}
\caption{ Comparison of our observations to extrapolated \fgas\ from scaling relations \citep{Tacconi2013,Scoville2017, Tacconi2018, Liu2019} as a function of sSFR at fixed $z=1.5$, for $10.8<$\Mstar$<11.6$ (indicated by shaded region) to match the range of observations plotted \citep[symbols with black edges; this work and][]{Gobat2018}.  \textsc{simba} quiescent galaxies that meet our selection criteria are circles (those with no ongoing SFR have a floor set to log$_{10}$ sSFR = -13 yr$^{-1}$).  Our observed limits are in agreement with the low gas fractions predicted by \textsc{simba} simulations, and both have significantly lower \fgas than expected from scaling relations. 
}
\end{center}
\end{figure*}

\subsection{Comparison to analytic bathtub models}

Further insight is possible by comparing to analytical ``bathtub'' models, where the gas content of galaxies is an equilibrium of gas infow, outflow and consumption by star formation \citep[e.g.][]{Dave2012, Finlator2008, Bouche2010, Lilly2013, PengMaiolino2014, RathausSternberg2016}. This self-regulation, 
to first order, appears to describe the behavior of \fgas\ across star forming galaxy populations remarkably well \citep{Tacconi2020}.
However, as halos grow above \Mhalo$>10^{12}$ \msol, the accretion of baryons is slowed down due to shock heating at the virial radius \citep[e.g.][]{DekelBirnboim2006}. More massive halos reach this critical mass at higher redshifts, spending a longer fraction of cosmic time without accreting new fuel for star formation.

In this section, we compare our observed gas fractions to that predicted using the simple analytic equilibrium model for \fgas(z, \Mhalo) outlined in \citet[][]{Dave2012}. For a given halo mass and formation redshift, gas in the galaxy is computed from cosmological accretion as a function of \Mhalo\ \citep{Dekel2009}, simple stellar and preventative feedback prescriptions that remove gas or keep it hot in the halo, and consumption from the star formation rate. Although the gas fraction in this model is not a self-consistent model for gas evolution because it is computed from the star formation rate with an assumed star formation efficiency, this comparison nonetheless is a simple intuitive tool to qualitatively compare the relative impact of competing processes in galaxies that affect the gas fraction evolution.

In Figure \ref{fig:fgas_dave} we show a series of these models in comparison to our observations. We show \fgas\ for galaxies in halos that reach masses at z=0 of \Mhalo$=$10$^{11}$ (black), 10$^{12}$ (magenta), 10$^{13}$ (orange) and 10$^{14}$ (yellow) \msol\ published in \citet[][]{Dave2012}. Observed galaxies are color-coded by their inferred halo mass at the redshift of observation, using the stellar mass to halo mass relation of \citealt{Behroozi2010} as implemented in \textsc{halotools} \citep[][assuming no scatter, therefore the uncertainties are likely large]{Hearin2017}.  

This model predicts that only halos that reach 10$^{14}$\msol\ by $z=0$ halt accretion early enough in cosmic time to allow gas consumption to reach the low \fgas\ we observe in our sample. A halo with 10$^{14}$\msol\ at $z=0$ reaches this critical halo mass of 10$^{12}$\msol\ at $z\sim4$, and exceeds the quenching threshold (which evolves slightly with z) around $z
\sim3$ \citep{Dekel2009, Dave2012}. For \Mhalo $\lesssim$10$^{13}$\msol\ there is not enough time to consume the gas already accreted, and other effects would be required (e.g. gas destruction from feedback) to match our low gas fractions. 
For Mhalo $>10^{13}$ we also plot additional mathematical forms to describe the outflow term in the equilibrium model (variations to the stellar feedback prescription, which vary the star formation efficiency). The dotted line indicates a mass loading factor that lowers efficiency at low masses, and the dashed line indicates an additional dependence on metallicity, that decreases gas consumption at low metallicity. These variations mostly impact growth and gas fraction at low galaxy mass and improve agreement with observations at low masses, but for our case the differences are small and do not impact this result.

Based on these models we speculate that a plausible explanation of our observations is that our galaxies reside in massive halos (10$^{14}$\msol\ by $z=0$) that grew above the critical mass of 10$^{12}$\msol\, slowing gas accretion early enough in cosmic time ($z\sim4$) to reach low gas fractions by $z\sim1.5$. This scenario is qualitatively similar to the idea of cosmological starvation explored in \citep{FeldmannMayer2015}.

Estimated halo masses for the ALMA sample are consistent with this picture. The stellar mass to halo mass relation predicts typical log$_{10}$\Mhalo/\msol\ for our sample of $13.5-14$ at their respective redshifts \citep{Behroozi2010}, in general agreement with inferred halo masses from clustering of quiescent galaxies at $z>1$ \citep[e.g.][]{Ji2018}. 
Furthermore, the relative number density of our ALMA sample from integrating the observed stellar mass function \citep[$\sim$10$^{-5}$Mpc$^{-3}$;][]{Tomczak2014} is similar to  that of log$_{10}$\Mhalo/\msol$>$13.5 at our typical redshift $z\sim1.4$ \citep[halos that will reach 10$^{14}$\msol\ by z=0; calculated using the halo mass function calculator \textsc{hmf} published by][assuming the halo mass function of \citealt{Behroozi2013}]{Murray2013}. 
Were our sample too numerous compared to the requisite mass halos, it would require some fraction of lower mass halos have their gas destroyed more rapidly than implied by the equilibrium model (e.g. via stronger AGN feedback). We note that these ballpark estimates are uncertain owing to scatter in the stellar mass to halo mass relation as well as uncertainties in linking progenitor populations through cumulative number density evolution \citep{Wellons2017, Torrey2017}.

Unfortunately, the simplicity of this analytical model and the significant intrinsic scatter in the stellar mass to halo mass relation precludes a rigorous test of the idea that reaching high halo mass and stopping accretion at early times is the primary driver of low gas fractions. We can only speculate here that this could be a contributing factor. With recent improvements in cosmological simulations, they may provide more realistic and self-consistent comparisons to observables like \fgas. We explore these comparisons in the next section.

\subsection{Comparison to cosmological simulations}

Historically, cosmological simulations have been challenged to match massive galaxies in their abundances over cosmic time, as well as to prevent continued star formation in massive quiescent galaxies \citep[for a review see][]{SomervilleDave2015}. Recent advances in feedback prescriptions have enabled progress on both of these fronts \citep{Vogelsberger2014, Schaye2015, Dave2019}, and now face a new challenge to match the ISM properties such as the cold gas reservoirs we study here \citep[e.g.][]{Narayanan2012b, Lagos2014, Lagos2015a, Lagos2015b}.  Analysis of recent cosmological hydrodynamical simulations indicate that modern implementations of feedback prescriptions for massive galaxies are able to qualitatively reproduce the global scaling relations for star forming galaxies across cosmic time \citep[e.g.][]{Scoville2017, Tacconi2018, Liu2019} as well as the low \fgas\ that are observed in massive and quiescent galaxies by $z\sim0$ \citep[e.g.][]{Young2011,Saintonge2017, Davis2016}. With our new observations of \fgas\ presented here we can now extend these comparisons to massive quiescent galaxies at $z\sim1.5$.  
We compare to the predictions for molecular gas reservoirs in the \textsc{simba} simulation \citep{Dave2019}. \textsc{simba} quenches galaxies primarily via its implementation of jet AGN feedback, in which $\sim10^4$ km/s jets are ejected bipolarly from low-Eddington ratio black holes.  The jets are explicitly decoupled from the ISM, thus presumably the quenching owes to heating and/or removal of halo gas.  \textsc{simba}'s X-ray feedback is important for removing H$_2$ from the central regions \citep[$<0.5R_{e}$;][]{Appleby2020}, which may also contribute to lowering the global molecular content, in general agreement with evidence for inside out quenching observed in molecular gas reservoirs \citep{Spilker2019}.

We select quiescent galaxies from a snapshot at $z\sim1.5$ to match our ALMA target selection criteria: log$_{10}$\Mstar/\msol$>$11.3 and log$_{10}$ sSFR$<-10$yr$^{-1}$. 
The comparison of the \fgas\ in \textsc{simba} galaxies  compared to our ALMA observations can be seen in Figure \ref{fig:fgas_simba}. Remarkably, \textsc{simba} predicts low \fgas\ in quiescent galaxies that are consistent with our observational limits.   Our ALMA limits on \fgas\ lie at the upper envelope of \fgas\ predicted for \textsc{simba} galaxies of similar mass and sSFR, with the majority of \textsc{simba} galaxies containing \fgas$<$3\%. 
90\% of \textsc{simba} galaxies similar to our sample reside in halos with log$_{10}$\Mhalo/\msol$>$13, and likely truncated accretion of new gas at earlier times. Better observational constraints on \tdep, the time evolution of gas reservoirs, and the precision of stellar age diagnostics would be required to link this success directly to the destruction from feedback model, and/or the truncation of new gas accretion as explored in the previous section. 
\textsc{simba} produces a comparable population of ``slow quenchers" and ``fast quenchers" \citep{RodriguezMontero2019} at these redshifts, and in the future we will examine whether the galaxies consistent with our ALMA limits are preferentially in either category, and measure the associated gas depletion times.

Also in Figure \ref{fig:fgas_simba} we show the scaling relations based on star forming galaxies across redshifts and quiescent galaxies at $z\sim0$ \citep{Scoville2017, Tacconi2018, Liu2019}.  The shaded regions correspond to the scaling relations at z$=1.5$ for log$_{10}$\Mstar/\msol=10.8 (upper bound set by the average mass of the sample studied in \citealt{Gobat2018}) and log$_{10}$\Mstar/\msol=11.6 (lower bound set by the mass of our most massive galaxy). Both our ALMA limits, as well as the \textsc{simba} predictions, lie well below scaling relations for $\fgas(\Mstar,z,sSFR)$. This is consistent with the results of Section \ref{sec:timescale}, indicating that the simulations also disagree with extrapolations of current scaling relations. 
Improvements to future scaling relations should include data from surveys such as this one, in the poorly explored parameter space of high redshift and low sSFR.

\section{Conclusions}
We have conducted the first molecular gas survey of massive quiescent galaxies at $z>1$, using CO(2--1) measured with ALMA. We summarize the findings of our survey as follows:

1. We find very low \fgas $<2-6$\% measured for massive quiescent galaxies at $z\sim1.5$. The sample uniformly displays \fgas $<6\%$ and we do not observe any variation with size or stellar density across the large dynamic range of the structural properties within our sample.

2. Depletion times must be rapid as galaxies leave the star forming sequence in order to match our constraints of very low \fgas. We estimate an upper limit to the typical depletion time of \tdep$<0.6$ Gyr, much shorter than expected from extrapolating current scaling relations to low sSFR.

3. Our low \fgas\ limits are generally consistent with the predictions of an analytical ``bathtub" model, for galaxies in massive halos that reach log$_{10}$\Mhalo/\msol=14 by z=0. 
We speculate that  ``cosmological starvation" after reaching a critical mass of log$_{10}$\Mhalo/\msol=12 ($z\sim4$ for these halos), contributed to the rapid decline in \fgas\ required by our observations.

4. Our low \fgas\ limits are consistent with predictions from the recent \textsc{simba} cosmological simulations with realistic AGN feedback, highlighting another success for state-of-the-art models describing the properties of massive quiescent galaxies. This consistency, like the bathtub model, may also point to the simple truncation and consumption picture. However, with our data we cannot rule out that low gas fractions result from gas destruction from feedback or an increase in the efficiency of gas consumption.

Although it may be observationally expensive, concrete tests of current and future galaxy formation models will rely on building larger datasets that probe the molecular gas properties of galaxies with little on-going star formation. Building statistical samples will be challenging and there are a number of approaches that one could take. Real progress will be made with increasing numbers alone. Another possibility would be to combine information about the SFHs with depletion time tracks to follow individual objects back in time. The extraction of these histories from quiescent galaxies at cosmic noon will soon be enabled by the unparalleled capabilities of the {\it James Webb Space Telescope} ({\it JWST}). Deep photometric and spectrosopic surveys are planned for Cycle 1 \citep{Williams2018, Rieke2019} that will be capable of identifying quiescent galaxies even at $z>4$ and reconstructing their star formation histories with unprecedented detail. These will make ideal targets for future ALMA CO surveys to build our understanding of molecular gas in galaxies that have ceased star formation.

\acknowledgments

This work was performed in part at Aspen Center for Physics, which is supported by National Science Foundation grant PHY-1607611. We acknowledge valuable discussions with Ivo Labbe, Wren Suess, Sandro Tacchella, Sirio Belli. CCW acknowledges support from the National Science Foundation Astronomy and Astrophysics Fellowship grant AST-1701546. JSS is supported by NASA Hubble Fellowship grant \#HF2-51446  awarded  by  the  Space  Telescope  Science  Institute,  which  is  operated  by  the  Association  of  Universities  for  Research  in  Astronomy,  Inc.,  for  NASA,  under  contract  NAS5-26555. K.E.W. wishes to acknowledge funding from the Alfred P. Sloan Foundation. CAW is supported by the National Science Foundation through the Graduate Research Fellowship Program funded by Grant Award No. DGE-1746060. This paper makes use of the following ALMA data: ADS/JAO.ALMA \#2018.1.01739.S, ADS/JAO.ALMA \#2015.1.00853.S. ALMA is a partnership of ESO (representing its member states), NSF (USA), and NINS (Japan), together with NRC (Canada), MOST and ASIAA (Taiwan), and KASI (Republic of Korea), in cooperation with the Republic of Chile. The Joint ALMA Observatory is operated by ESO, AUI/NRAO, and NAOJ. The National Radio Astronomy Observatory is a facility of the National Science Foundation operated under cooperative agreement by Associated Universities, Inc. The Cosmic Dawn Center is funded by the Danish National Research Foundation.

\bibliographystyle{aasjournal}
\bibliography{nondetections_paper}

\end{document}